\documentclass[times,twocolumn,final]{elsarticle}

\usepackage{lineno,hyperref}
\modulolinenumbers[5]

\journal{NeuroInformatics}

\usepackage{svg}
\usepackage{times}
\usepackage{epsfig}
\usepackage{graphicx}
\usepackage{float}
\usepackage{amsmath}
\usepackage{amssymb}
\usepackage{siunitx} 
\usepackage{bbm}
\usepackage{xcolor}
\usepackage{amsmath} 
\usepackage{multirow,booktabs}
\usepackage{mathtools}
\usepackage{soul}
\usepackage{xcolor}
\usepackage{url}
\usepackage[numbers]{natbib}

\mathchardef\mhyphen="2D 
  
\definecolor{red}{rgb}{0.234, 0.6835, 0.808}

\renewcommand{\hl}[1]{#1}

\begin{document}

\begin{frontmatter}

\title{Automatic cerebral hemisphere segmentation in \hl{rat} MRI with lesions via attention-based convolutional neural networks}

\author{Juan Miguel Valverde\textsuperscript{a}, Artem Shatillo\textsuperscript{b}, Riccardo de Feo\textsuperscript{a}, Jussi Tohka\textsuperscript{a}}
\address{\textsuperscript{a}A.I. Virtanen Institute for Molecular Sciences, University of Eastern Finland, 70150 Kuopio, Finland}
\address{\textsuperscript{b}Charles River Discovery Services, Kuopio 70210, Finland.}

\begin{abstract}
We present MedicDeepLabv3+, a convolutional neural network that is the first completely automatic method to segment cerebral hemispheres in magnetic resonance (MR) volumes of \hl{rats} with lesions.
MedicDeepLabv3+ improves the state-of-the-art DeepLabv3+ with an advanced decoder, incorporating spatial attention layers and additional skip connections that, as we show in our experiments, lead to more precise segmentations.
MedicDeepLabv3+ requires no MR image preprocessing, such as bias-field correction or registration to a template, produces segmentations in less than a second, and its GPU memory requirements can be adjusted based on the available resources.
We optimized MedicDeepLabv3+ and \hl{six} other state-of-the-art convolutional neural networks (DeepLabv3+, UNet, \hl{HighRes3DNet, V-Net, VoxResNet}, Demon) on a heterogeneous training set comprised by MR volumes from 11 cohorts acquired at different lesion stages.
Then, we evaluated the trained models and two approaches specifically designed for rodent MRI skull stripping (RATS and RBET) on a large dataset of 655 MR rat brain volumes.
In our experiments, MedicDeepLabv3+ outperformed the other methods, yielding an average Dice coefficient of 0.952 and 0.944 in the brain and contralateral hemisphere regions.
Additionally, we show that despite limiting the GPU memory and the training data, our MedicDeepLabv3+ also provided satisfactory segmentations.
In conclusion, our method, publicly available at \url{https://github.com/jmlipman/MedicDeepLabv3Plus}, yielded excellent results in multiple scenarios, demonstrating its capability to reduce human workload in \hl{rat} neuroimaging studies.
\end{abstract}

\end{frontmatter}

\section{Introduction}
Rodents are widely used in preclinical research to investigate brain diseases \citep{carbone2021estimating}.
These studies often utilize in-vivo imaging technologies, such as magnetic resonance imaging (MRI), to visualize brain tissue at different time-points, which is necessary for studying disease progression.
MRI permits the acquisition of brain images with different contrasts in a non-invasive manner, making MRI a particularly advantageous in-vivo imaging technology.
However, these images typically need to be segmented before conducting quantitative analysis.
As an example, the size of the hemispheric brain edema relative to the brain size is an important biomarker for acute stroke that requires accurate hemisphere segmentation \citep{swanson1990semiautomated,gerriets2004noninvasive}.

With brain edema biomarkers in mind, our work focuses on cerebral hemisphere segmentation in \hl{rat} MR images with lesions.
Segmenting these images is particularly challenging since lesions' size, shape, location, and contrast can vary even within images from the same cohort, hampering, as we show in our experiments, traditional segmentation methods. In addition, rodents' small size makes image acquisition sensitive to misalignments, potentially producing slices with asymmetric hemispheres and particularly affecting anisotropic data.
These difficulties have led researchers and technicians to annotate rodent cerebral hemispheres manually \citep{freret2006long,mcbride2015correcting}, which is laborious and time-consuming, and motivates this work.

In recent years, convolutional neural networks (ConvNets) have been widely used to segment medical images due to their outstanding performance \citep{bakas2018identifying,bernard2018deep,heller2021state}.
ConvNets can be optimized end-to-end, require no preprocessing, such as bias-field correction and costly registration, and can produce segmentation masks in real time \citep{de2021automated}.
ConvNets can also be tailored to specific segmentation problems by incorporating domain constraints and shape priors \citep{kervadec2019constrained}.
In particular, DeepLabv3+ architecture with its efficient computation of large image regions via dilated convolutions has demonstrated excellent results on various segmentation tasks \citep{chen2018encoder}, leading researchers to its application on medical image segmentation.
\citet{xie2019deep} utilized DeepLabv3+ to estimate the segmentation maps and subsequently refined such estimation with a second ConvNet.
\citet{ma2019neural} modified DeepLabv3+ for applying style transfer to homogenize MR images with different properties.
\citet{khan2020evaluation} showed that DeepLabv3+ outperforms other ConvNets on prostate segmentation of T2-weighted MR scans.
However, in our preliminary experiments, DeepLabv3+ provided unsatisfactory results, especially in the masks borders.

We present and make publicly available MedicDeepLabv3+, the first method for segmenting cerebral hemispheres in MR images of \hl{rats} with lesions.
MedicDeepLabv3+ improves DeepLabv3+ architecture with a new decoder with spatial attention layers \citep{oktay2018attention,wang2019automatic} and an increased number of skip connections that facilitate the optimization.
We optimized our method on a training set comprised by 51 MR rat brain volumes from 11 cohorts acquired at multiple lesion stages, and we evaluated it on a large and challenging dataset of 655 MR rat brain volumes.
Our experiments show that MedicDeepLabv3+ outperformed the baseline state-of-the-art DeepLabv3+ \citep{chen2018encoder}, UNet \citep{ronneberger2015u}, \hl{HighRes3DNet} \citep{li2017compactness}, \hl{V-Net} \citep{milletari2016v}, \hl{VoxResNet} \citep{chen2018voxresnet}, and, particularly for skull stripping, it also outperformed Demon \citep{roy2018deep}, RATS \citep{oguz2014rats}, and RBET \citep{wood2013rbet}.
Additionally, we evaluated MedicDeepLabv3+ with very limited GPU memory and training data, and our experiments demonstrate that, despite such restrictions, MedicDeepLabv3+ yields satisfactory segmentations, showcasing its usability in multiple real-life situations and environments.

\subsection{Related work}
\paragraph{Anatomical segmentation of rodent brain MRI with lesions.} Anatomical segmentation in MR images of rodents with lesions is an under-researched area; \citet{roy2018deep} and \citet{riccardohippocampus} are the only studies that examined this problem.
\citet{roy2018deep} showed that their Inception-based \citep{szegedy2015going} skull-stripping ConvNet named `Demon' outperformed other methods on MR images of mice and humans with traumatic brain injury.
\citet{riccardohippocampus} presented an ensemble of ConvNets named MU-Net-R for ipsi- and contralateral hippocampus segmentation on MR images of rats with traumatic brain injury.
\citet{mulder2017automated} developed a lesion segmentation pipeline that includes an atlas-based contralateral hemisphere segmentation step. However, these hemisphere segmentations were not compared to a ground truth, and this approach is sensitive to the lesion appearance because it relies on registration.

\paragraph{Anatomical segmentation of rodent brain MRI without lesions.} The vast majority of anatomical segmentation methods for rodent MR brain images have been exclusively developed for brains without lesions. These methods can be classified into three categories. First, atlas-based segmentation approaches, which apply registration to one or more brain atlases \citep{pagani2016semi} and, afterwards, label candidates are refined or combined with, for instance, Markov random fields \citep{ma2014automatic}. As these approaches heavily rely on registration, they underperform in the presence of anatomical deformations. Second, methods that group nearby voxels with similar properties. These approaches typically start by proposing one or several candidate regions, and later adjust such regions with an energy function and, optionally, shape priors. Examples of these methods include surface deformation models \citep{wood2013rbet}, graph-based segmentation algorithms  \citep{oguz2014rats}, and a more recent approach that combines blobs into a single region \citep{liu2020automatic}. These approaches can handle different MRI contrasts and require no registration. However, they also rely on local features, such as nearby image gradients and intensities. Thus, these methods can be very sensitive to intensity inhomogeneities, and small brain deformities.
Third, machine learning algorithms that classify brain features.
These features can be handcrafted, such as in \citep{bae2009automated,wu2012prior} where authors employed support vector machines to classify voxels into different neuroanatomical regions based on their intensity, location, neighbor labels, and probability maps.
On the contrary, deep neural networks, a subclass of machine learning algorithms, can automatically find relevant features and learn meaningful non-linear relationships between such features.
Methods based on neural networks, such as pulse-coupled neural networks \citep{chou2011robust,murugavel2009automatic} and ConvNets \citep{roy2018deep,hsu2020automatic,de2021automated}, have been used in the context of rodent MRI segmentation.

\paragraph{Lesion segmentation of rodent brain MRI.} The high contrast between lesion and non-lesion voxels in certain rodent brain MR images motivated the development of thresholding-based methods \citep{wang2007comparing,choi2018novel}.
However, these methods are not fully automatic, and they cannot be used in MR images with other contrasts, or lesions with different appearances.
\citet{mulder2017automated} introduced a fully-automated pipeline to segment lesions via level sets \citep{dervieux1980finite,osher1988fronts}.
Images were first registered to a template, then skull stripped, and their ventricles were segmented prior to the final lesion segmentation step.
\citet{arnaud2018fully} framed lesion segmentation as an anomaly-detection problem and developed a pipeline that detects voxels with unusual intensity values with respect to healthy rodent brains.
\citet{valverde2020ratlesnetv2} developed the first single-step method to segment rodent brain MRI lesions using ConvNets.

\section{Materials and methods}
\subsection{MRI Data}
\begin{figure}[t]
\begin{center}
   \includegraphics[width=0.98\linewidth]{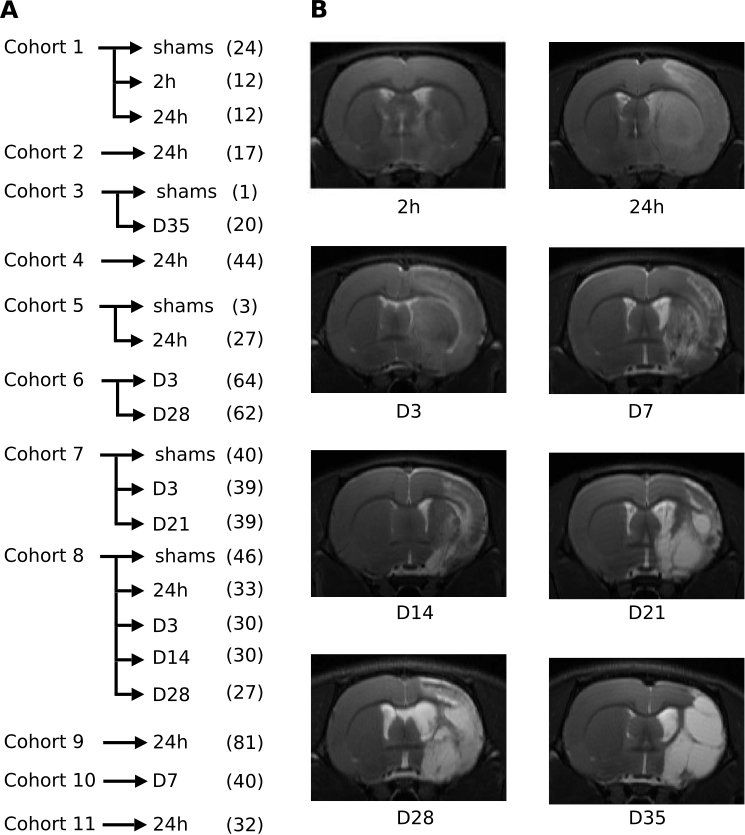}
\end{center}
   \caption{A: Cohorts, acquisition time-points, and number of images. B: Example slice from each lesion stage in approximately the same brain area.}
\label{fig:data}
\end{figure}

The image data, provided by Charles River Laboratories Discovery site (Kuopio, Finland)\footnote{https://www.criver.com/products-services/discovery-services}, consisted of 723 MR T2-weighted brain scans of 481 adult male Wistar rats weighting between 250-300 g derived from 11 different cohorts. Rats were induced focal cerebral ischemia by middle cerebral artery occlusion for 120 minutes in the right hemisphere of the brain \citep{koizumimodel}.
MR data was acquired at multiple time-points after the occlusion; for each of the 11 cohorts, time-points were different (see Fig. \ref{fig:data}-A for details).
In total, our dataset contained MR images from nine lesion stages: shams, 2h, 24h, D3, D7, D14, D21, D28, and D35.
Figure \ref{fig:data}-B shows representative images of these lesion stages in approximately the same brain area.
All animal experiments were conducted according to the National Institute of Health (NIH) guidelines for the care and use of laboratory animals, and approved by the National Animal Experiment Board, Finland. Multi-slice multi-echo sequence was used with the following parameters; TR = 2.5 s, 12 echo times (10-120 ms in 10 ms steps) and 4 averages in a horizontal 7T magnet. T2-weighted images were calculated as the sum of the all echoes. Eighteen coronal slices of 1 mm thickness were acquired using a field-of-view of 30x30 mm$^2$ producing 256x256 imaging matrices of resolution $117 \times 117$ \SI{}{\micro\meter}.

\subsection{Data preparation}
The T2-weighted images were not preprocessed (i.e., no registration, bias-field or artifact correction), and their intensity values were standardized to have zero mean and unit variance.
Brain and contralateral hemisphere masks were annotated by several trained technicians employed by Charles River according to a standard operating procedure.
These annotations did not include the cerebellum and the olfactory bulb.
Finally, we computed the ipsilateral hemisphere mask by subtracting the contralateral hemisphere from the brain mask, yielding non-overlapping regions (i.e., the background, ipsilateral and contralateral hemispheres) for optimizing the ConvNets.

\subsection{Train, validation and test sets} \label{sec:dataprep}
We divided the MR images into a training set of 51 volumes, validation set of 17 volumes, and test set of 655 volumes.
Specifically, we grouped the MR images by their cohort and acquisition time-point (Fig. \ref{fig:data}-A).
From the resulting 17 subgroups, our training and validation sets comprised 3 and 1 MR images, respectively, per subgroup.
Images from sham-operated animals were not included to the training and validation sets since our work focused on \hl{rat} brains with lesions.
The remaining 655 MR images, including shams, formed the independent test set.
This splitting strategy aimed to create a diverse training set, as brain lesions have notably different T2-weighted MRI intensities depending on the lesion stage, and annotations can differ slightly across cohorts due to the task subjectivity and the consequent low inter-rater agreement \citep{mulder2017automated,valverde2020ratlesnetv2}.

\begin{figure*}[t]
\centering
\includegraphics[width=0.9\textwidth]{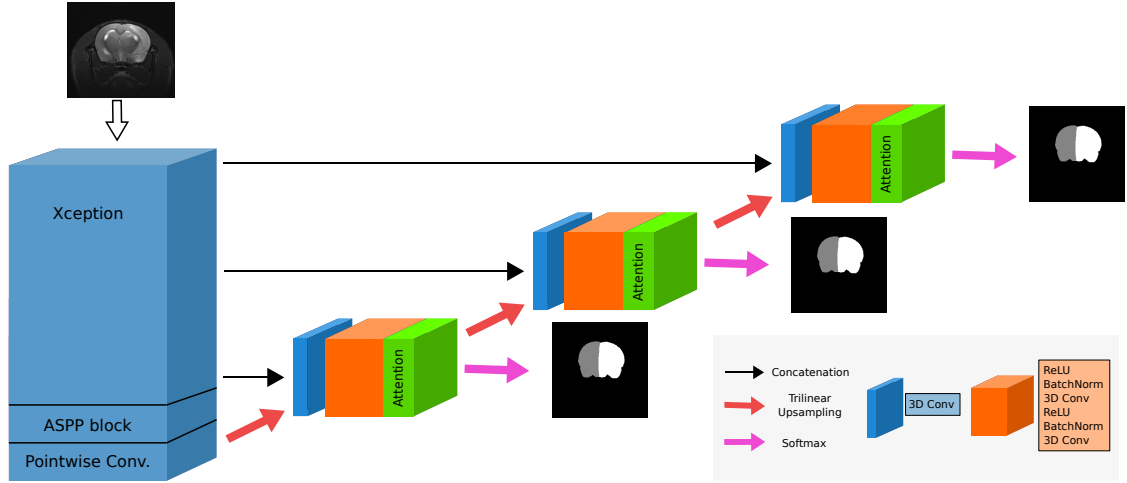}
\caption{MedicDeepLabv3+ architecture.}
\label{fig:architecture}
\end{figure*}

\subsection{MedicDeepLabv3+}
MedicDeepLabv3+ is a 3D fully convolutional neural network (ConvNet) based on DeepLabv3+ \citep{chen2018encoder} and UNet \citep{ronneberger2015u}.
\hl{We chose DeepLabv3+ because of its excellent performance in semantic segmentation tasks, and we modified its last layers to resemble more closely to UNet, which is an architecture widely used in medical image segmentation.}
DeepLabv3+ first employs Xception \citep{chollet2017xception} to transform the input and reduce its dimensionality, and then it upsamples the transformed data, twice, by a factor of four.
MedicDeepLabv3+ replaces these last layers (i.e., the decoder) with three stages of skip connections and convolutional layers, and, as we describe below, it incorporates custom spatial attention layers, enabling deep supervision (see Fig. \ref{fig:architecture}).

\subsubsection{Encoder}
MedicDeepLabv3+ stacks several $3 \times 3 \times 3$ convolutional layers, normalizes the data with Batch Normalization \citep{ioffe2015batch}, and incorporates residual connections \citep{he2016deep}.
\hl{Both batch normalization and residual connections are well established architectural components that have been shown to facilitate the optimization in deep ConvNets} \citep{drozdzal2016importance,li2018visualizing}.
The first layers of MedicDeepLabv3+ correspond to Xception \citep{chollet2017xception}, which uses depthwise-separable convolutions instead of regular convolutions.
\hl{These depthwise-separable convolutions are advantageous over regular convolutions as they can decouple channel and spatial information.
This is achieved by separating the operations of a regular convolution into a spatial feature learning and a channel combination step, increasing the efficiency and performance of the model }\citep{chollet2017xception}.

MedicDeepLabv3+ utilizes dilated convolutions in the last layer of Xception.
Dilated convolutions sample padded input patches and multiply the non-padded values with the convolution kernel, thus, expanding the receptive field of the network \citep{chen2014semantic}.
\hl{In other words, dilated convolutions permit to adjust the area that influences the classification of each voxel, and, as increasing this area has shown to improve model performance, we opted to employ dilated convolutions as in DeepLabv3+} \citep{chen2017deeplab}.
After the Xception backbone, DeepLabv3+'s Atrous Spatial Pyramid Pooling (ASPP) module concatenates parallel branches of dilated convolutional layers with different dilation rates and an average pooling followed by trilinear interpolation.
Then, a pointwise convolution combines and reduces the number of channels.
To this step, the described architecture reduces the data dimensionality by a factor of 16.

\subsubsection{Decoder} \label{sec:decoder}
We developed a new decoder for MedicDeepLabv3+ with more stages of skip connections and convolution blocks than DeepLabv3+.
\hl{In each stage, feature maps are upsampled via trilinear interpolation and concatenated to previous feature maps from the encoder.
Subsequently,} $3 \times 3 \times 3$ convolutions halve the number of channels (Figure \ref{fig:architecture}, blue blocks), and a ResNet block \citep{he2016deep} further transforms the data (Figure \ref{fig:architecture}, orange blocks).
The consequent increase of skip-connections facilitates MedicDeepLabv3+ optimization \citep{drozdzal2016importance,li2018visualizing}.
Importantly, DeepLabv3+ produces segmentations at $\times 4$ less resolution than the original images that, to match their size, are upsampled via interpolation.
In contrast, our MedicDeepLabv3+ incorporates more convolutional layers at the end of its architecture to directly produce segmentations at the original size.

\begin{figure}[t]
\centering
\includegraphics[width=0.98\linewidth]{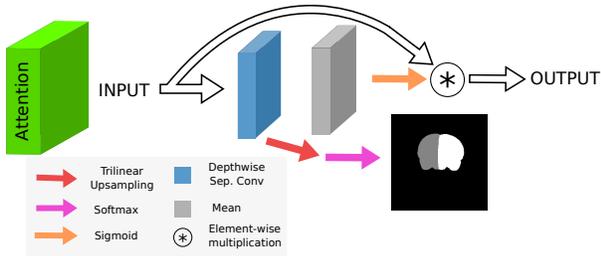}
\caption{Spatial attention block (details in Section \ref{sec:decoder}).}
\label{fig:attentionlayer}
\end{figure}

Another key difference with respect to DeepLabv3+ is that MedicDeepLabv3+ utilizes spatial attention layers that learn and apply voxel-wise importance maps \citep{oktay2018attention,wang2019automatic}.
\hl{Attention layers have been shown to aid the network to identify and focus on the most important features, improving performance }\citep{fu2019dual}\hl{.
In our implementation, these} layers (Fig. \ref{fig:attentionlayer}) transform the inputs with a depthwise-separable convolution and, subsequently, average the resulting feature maps.
Afterwards, a sigmoid activation function transforms the data non-linearly, producing spatial attention maps with values in the range $[0, 1]$. Then, these attention maps multiply the input feature maps voxel-wise.
To encourage spatial attention maps that lead to the ground truth and to further facilitate the optimization, we added a branch in the first two attention layers for generating downsampled probability maps of the segmentation masks, enabling deep supervision (Fig \ref{fig:attentionlayer}, red and pink arrows).

\subsubsection{Loss function} \label{sec:loss}
We trained MedicDeepLabv3+ with deep supervision \citep{lee2015deeply}, i.e., we minimized the sum of cross entropy and Dice loss of all outputs of MedicDeepLabv3+ (Fig \ref{fig:architecture}, pink arrow).
Formally, we minimized $L = \sum_{s \in S} L_{CE}^s + L_{Dice}^s$ with $S = \{1,2,3\}$ indicating each MedicDeepLabv3+ output (see Fig. \ref{fig:architecture}).
Cross entropy treats the model predictions and the ground truth as distributions

\begin{equation}
    \label{eq:bce}
    L_{CE}=-\frac{1}{NC} \sum_{i=1}^{N} \sum_{c=1}^{C} p_{i,c} \log(q_{i,c}),
\end{equation}
where $p_{i,c} \in \{0, 1\}$ represents whether voxel $i$ belongs to class $c$, and $q_{i,c} \in [0, 1]$ its predicted Softmax probability. $C = 3$ for background, ipsilateral and contralateral hemisphere classes, and $N$ is the total number of voxels. Dice loss estimates the Dice coefficient between the predictions and the ground truth:

\begin{equation}
    \label{eq:diceloss}
    L_{Dice}=1 - \frac{2}{C} \sum_{c=1}^C \frac{\sum_{i}^{N} p_{i,c} q_{i,c}}{\sum_{i}^{N} p_{i,c}^{2}+ q_{i,c}^{2}}.
\end{equation}

\hl{Minimizing cross entropy and Dice loss is a common practice in medical image segmentation }\citep{myronenko2019robust,isensee2021nnu}\hl{.
Cross entropy optimization reduces the difference between the ground truth and prediction distributions.
Dice loss optimization increases the Dice coefficient that we ultimately aim to maximize and it is particularly beneficial in class-imbalanced datasets} \citep{milletari2016v}.
Additionally, their optimization at different stages via deep supervision is equivalent to adding shortcut connections to propagate the gradients to various layers, facilitating the optimization of those layers.

\subsection{Experimental design}
\subsubsection{Metrics}
We assessed the automatic segmentations with Dice coefficient \citep{dice1945measures}, Hausdorff distance \citep{rote1991computing}, precision, and recall. Dice coefficient measures the overlapping volume between the ground truth and the prediction
\begin{equation}
    \label{eq:dicecoeff}
    Dice(A, B) = \frac{2|A \cap B|}{|A| + |B|},
\end{equation}
where $A$ and $B$ are the segmentation masks. Hausdorff distance (HD) is a quality metric that calculates the distance to the misclassification located the farthest from the boundary masks. Formally:

\begin{equation}
d(A, B)=\max \left\{\max _{a \in \partial A} \min _{b \in \partial B}|b-a|, \max _{b \in \partial B} \min _{a \in \partial A}|a-b|\right\},
\end{equation}
where $\partial A$ and $\partial B$ are the boundary voxels of $A$ and $B$, respectively. In other words, HD provides the distance to the largest segmentation error. We provided HD values in mm and we accounted for voxel anisotropy.
Finally, precision is the percentage of voxels accurately classified as brain/hemisphere, and recall is the percentage of brain/hemisphere voxels that were correctly identified:

\begin{equation}
    Prec = \frac{TP}{TP+FP} \hspace{0.4in} Recall = \frac{TP}{TP+FN}.
\end{equation}

\subsubsection{Benchmarked methods}
We compared our MedicDeepLabv3+ with DeepLabv3+ baseline \citep{chen2018encoder}, UNet \citep{ronneberger2015u}, \hl{HighRes3DNet} \citep{li2017compactness}, \hl{V-Net} \citep{milletari2016v}, \hl{VoxResNet} \citep{chen2018voxresnet}, Demon \citep{roy2018deep}, RATS \citep{oguz2014rats}, and RBET \citep{wood2013rbet}.
Since Demon, RATS, and RBET were exclusively designed for rodent skull stripping, we computed contralateral hemisphere masks only with MedicDeepLabv3+, DeepLabv3+, UNet, \hl{HighRes3DNet, V-Net, and VoxResNet}.
MedicDeepLabv3+ \hl{and all the other ConvNets} were optimized with Adam \citep{Kingma2014AdamAM} ($\beta_1 = 0.9$ ,$\beta_2 = 0.999$, $\epsilon = 10^{-8}$), starting with a learning rate of $10^{-5}$.
MedicDeepLabv3+, DeepLabv3+, \hl{HighRes3DNet, V-Net, VoxResNet} and UNet were trained for 300 epochs, and Demon was trained for an equivalent amount of time.
We ensembled three models \citep{dietterich2000ensemble} since this strategy markedly improved segmentation performance in our previous work \citep{valverde2020ratlesnetv2}.
More specifically, we trained each ConvNet three times, separately, starting from different random initializations.
Then, we formed the final segmentations based on the majority vote from the \hl{binarized} outputs of the three trained models.

\hl{We conducted a grid-search for best hyperparameters for RATS and RBET.
We performed the grid-search using merged training and validation sets as RATS and RBET do not involve supervised learning, thus making it possible to use also the training set for hyper-parameter tuning.}
Subsequently, we utilized the best-performing hyper-parameters on the test set.
With RATS, computing the brain mask $\hat{y}$ of image $x$ requires setting three hyper-parameters: intensity threshold $t$, $\alpha$, and rodent brain volume $s$, i.e., $\hat{y} = RATS(x, t, \alpha, s)$.
As rat brain volumes are highly similar in adult rats, we left this hyper-parameter with its default value, $s=1650$.
Thus, we only optimized for the threshold $t$ and $\alpha$ hyper-parameters.
Since RATS assumes that all intensity values are positive integers, we employed unnormalized images with RATS.
We optimized RATS hyper-parameters by maximizing the Dice coefficients in the training and validation sets:

\begin{equation}
    \underset{i, \alpha}{\arg \max } \hspace{0.07in} \sum_{x \in X_{train+val}} Dice(y, RATS(x, t, \alpha, 1650)) : t = P_{\%i},
\end{equation}
where $Dice$ is the Dice coefficient (Eq. (\ref{eq:dicecoeff})) between the ground-truth brain mask $y$ and RATS' output, $\alpha = 0, 1, \ldots, 10$ balances the importance between gradients and intensity values, and $P_{\%i}$ is the $i$th percentile of $x$ with $i = 0.01, 0.02, \ldots, 0.99$.
Since finding $t$ is potentially suboptimal due to the distribution variability across images, we optimized for the $i$th percentile, yielding image-specific thresholds.
In total, our hyper-parameter grid search in RATS comprised 1089 different parameter value combinations.
For RBET, we optimized the Dice coefficient to find the optimal ellipse axes ratio $w$:$h$:$d$ with $w,h,d$ from 0.1 to 1 in steps of 0.05, accounting for $19^3$ different configurations.
\hl{Note that, despite optimizing over a large number of hyper-parameter choices may increase the risk to overfit, our train, validation and test sets were derived so that $X_{train+val}$ is a good representation of $X_{test}$ (see Sec. }\ref{sec:dataprep}\hl{)}

Unlike ConvNets that can be optimized to segment specific brain regions, RATS and RBET perform skull stripping, segmenting also the cerebellum and olfactory bulb that were not annotated.
As these brain areas were not part of our ground truth, RATS and RBET segmentations would be unnecessarily penalized in those areas.
Thus, before computing the metrics, we discarded the slices containing cerebellum and olfactory bulb.
This evaluation strategy ignores potential misclassifications in the excluded slices, slightly favoring RATS and RBET.

\subsubsection{Brain midline evaluation} \label{sec:midline}
We calculated the average Dice coefficients of contra- and ipsilateral hemispheres around the brain midline---boundary between both hemispheres (see Fig. \ref{fig:brainmidline}-A).
Specifically, we considered the volume after expanding brain midline voxels in the coronal plane via morphological dilation $n$ times, with $n = 1, 2, \ldots, 10$.
In contrast to brain vs. non-brain tissue boundaries, the brain midline volume is more ambiguous to annotate due to the lower intensity contrast between hemispheres, hence the importance to assess the performance in this area.
\hl{This evaluation aims to supplement computing Dice coefficient and HD on the whole 3D volumes.
Since most of the voxels lie within the hemisphere borders, Dice coefficients tend to be very high, and since HD might indicate the distance to a misclassification that can be easily corrected via postprocessing (e.g., a misclassification outside the brain), HD alone does not suffice to assess specific areas.}
Note that, similarly to RATS and RBET evaluation, this experiment computed the Dice coefficient only on the slices that were manually annotated, as finding the brain midline requires these manual annotations.
Consequently, the evaluated \hl{3D} masks excluded non-annotated slices that could have false positives.

\subsubsection{Biomarkers based on hemisphere segmentation}
Since the ratio between contra- and ipsilateral hemispheres volume is an important biomarker for acute stroke \citep{swanson1990semiautomated,gerriets2004noninvasive}, we compared the hemisphere volume ratio of the ground truth with the hemisphere volume ratio of the automatic segmentations.
For this, we computed the effect size via Cohen's d \citep{lakens2013calculating} and the bias-corrected and accelerated (BCa) bootstrap confidence intervals \citep{efron1987better} with 100000 bootstrap resamples.
An effect size close to zero with a narrow confidence interval indicates a high similarity between automated and manual segmentation based biomarkers. 

\subsubsection{Performance with limited GPU memory and data}
Motivated by potential GPU memory limitations, we studied the performance and computational requirements of multiple versions of MedicDeepLabv3+ with lower capacity and, consequently, lower GPU memory usage.
To investigate this, we varied the number of kernel filters in all convolutions of MedicDeepLabv3+ that determines the number of parameters.
For instance, decreasing the number of kernel filters by half in the encoder also decreases the number of kernel filters in the decoder to half.

Separately, we evaluated the proposed MedicDeepLabv3+ on each cohort and time-point independently, simulating the typical scenario in rodent studies with extremely scarce annotated data.
For each of the 17 groups containing no sham animals (Fig. \ref{fig:data}-A), we trained an ensemble of three MedicDeepLabv3+ on \textit{only three images}, employed another image for validation during the optimization, and we evaluated this ensemble on the remaining holdout images from the same group.

\begin{figure*}[t]
\centering
\includegraphics[width=0.95\linewidth]{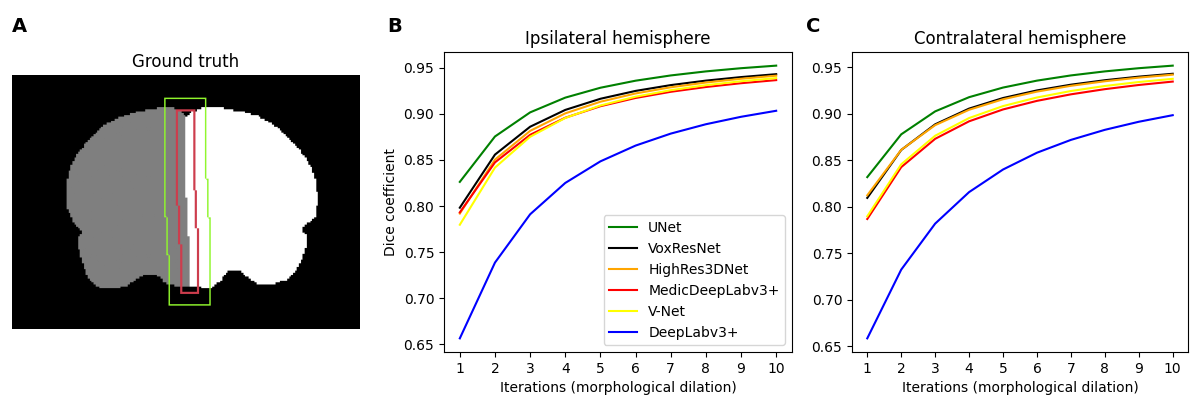}
\caption{A: Example of ground truth and its brain midline area after four (red) and ten (green) iterations of morphological dilation. B-C: Dice coefficients for the ipsi- and contralateral hemisphere classes in the brain midline area with different morphological dilation iterations (brain midline area sizes).}
\label{fig:brainmidline}
\end{figure*}

\subsubsection{Implementation} \label{sec:implementation}
MedicDeepLabv3+, DeepLabv3+, UNet, \hl{HighRes3DNet, V-Net, VoxResNet,} and Demon were implemented in Pytorch \citep{paszke2019pytorch} and were run on Ubuntu 16.04 with an Intel Xeon W-2125 CPU @ 4.00GHz processor, 64 GB of memory and an NVidia GeForce GTX 1080 Ti with 11 GB of memory.
MedicDeepLabv3+ and the scripts for segmenting \hl{rat} MR images and to optimize new models are publicly available at \url{https://github.com/jmlipman/MedicDeepLabv3Plus}.
These scripts are ready for use via command line interface with a single command, and users can easily adjust the number of initial filters that controls the model size, capacity, and GPU memory requirements.
Additionally, we provide the optimized parameters (i.e., the weights) of MedicDeepLabv3+ at \url{https://github.com/jmlipman/MedicDeepLabv3Plus}.

\section{Results}
\subsection{Segmentation metrics comparison}
Our MedicDeepLabv3+ produced brain and hemisphere masks with the highest Dice coefficients ($0.952$ and $0.944$) and precision ($0.94$ and $0.94$), and the lowest HD ($1.856$ and $2.064$) (see Table \ref{table:segmentationcomparison}).
\hl{MedicDeepLabv3+ also achieved the highest Dice coefficients in the brain and contralateral hemisphere most frequently, in 38\% and 36\% of the test images, respectively, followed by UNet (24\% and 26\%), VNet (13\% and 13\%), VoxResNet (13\% and 12\%), HighRes3DNet (11\% and 12\%) and the others (1\% or less)}.
All ConvNets performed better than RATS and RBET and, particularly, 3D ConvNets (MedicDeepLabv3+, DeepLabv3+, \hl{HighRes3DNet, V-Net, and VoxResNet}) consistently yielded lower HD than 2D ConvNets (UNet, Demon).
Our MedicDeepLabv3+ produced finer segmentations that were more similar to the ground truth than the baseline DeepLabv3+ which generated masks with imprecise borders.
UNet also produced segmentations with higher Dice and recall than DeepLabv3+, although UNet HD was considerably lower.
\hl{HighRes3DNet, V-Net, and VoxResNet yielded slightly worse Dice coefficients and HDs than MedicDeepLabv3+.}
Figure \ref{fig:segmentations} illustrates these results on the MR image with the highest hemispheric volume imbalance.
Figure \ref{fig:segmentations} shows that RBET was incapable \hl{of finding} the brain boundaries; RATS produced segmentations with several holes and non-smooth borders; 2D ConvNets misclassified the olfactory bulb and cerebellum; and, in agreement with Table \ref{table:segmentationcomparison}, MedicDeepLabv3+ produced the segmentation mask most similar to the ground truth.
We included 17 images (one per cohort and lesion time-point) in Online Resource 1 that also corroborate the higher performance of MedicDeepLabv3+.
The computation time to optimize these methods also varied notably: on average, ConvNets required 16 hours, and RATS and RBET needed six days.
Furthermore, MedicDeepLabv3+ segmented the images in real time, requiring approximately 0.4 seconds per image.

\subsection{Brain midline experiment}
Regarding the brain midline area experiment (Section \ref{sec:midline}, Figure \ref{fig:brainmidline}-B,C), MedicDeepLabv3+ outperformed the baseline DeepLabv3+ across different area sizes (average Dice coefficient difference of 0.07).
\hl{VoxResNet, HighRes3DNet, V-Net, and MedicDeepLabv3+ yielded very similar Dice coefficients}, and UNet produced the highest Dice coefficients by a small margin (average difference between UNet and MedicDeepLabv3+ of only 0.02).
Additionally, Dice coefficients were similar across hemispheres regardless of the segmentation method.

\begin{table*}
\begin{center}
\caption{Dice coefficients, Hausdorff distances (HD), precision, and recall of the brain and contralateral hemisphere (CH) masks derived from the evaluated methods \hl{(mean $\pm$ std)}. Bold: best scores.}
\label{table:segmentationcomparison}
\begin{tabular}{clllll}
\hline\noalign{\smallskip}
 & Approach & Dice & HD & Prec & Recall \\
\noalign{\smallskip}\hline\noalign{\smallskip}
\parbox[t]{2mm}{\multirow{9}{*}{\rotatebox[origin=c]{90}{Brain}}}
& MedicDeepLabv3+ & \textbf{0.952 $\pm$ 0.04} & \hl{\textbf{1.856 $\pm$ 0.91}} & 0.94 \hl{$\pm$ 0.07} & 0.97 \hl{$\pm$ 0.03} \\
& \hl{VoxResNet} & \hl{0.951 $\pm$ 0.04} & \hl{2.042 $\pm$ 1.02} & 0.94 \hl{$\pm$ 0.07} & 0.97 \hl{$\pm$ 0.02} \\
& \hl{HighRes3DNet} & \hl{0.949 $\pm$ 0.04} & \hl{1.858 $\pm$ 1.04} & 0.93 \hl{$\pm$ 0.07} & 0.97 \hl{$\pm$ 0.02} \\
& \hl{V-Net} & \hl{0.948 $\pm$ 0.04} & \hl{1.920 $\pm$ 1.05} & 0.94 \hl{$\pm$ 0.07} & 0.97 \hl{$\pm$ 0.02} \\
& UNet & 0.947 $\pm$ 0.05 & \hl{3.477 $\pm$ 1.20} & 0.93 \hl{$\pm$ 0.07} & 0.97 \hl{$\pm$ 0.02} \\
& DeepLabv3+ & 0.936 $\pm$ 0.04 & \hl{2.149 $\pm$ 1.02} & 0.93 \hl{$\pm$ 0.07} & 0.95 \hl{$\pm$ 0.03} \\
& Demon & 0.934 $\pm$ 0.04 & \hl{3.621 $\pm$ 1.17} & 0.92 \hl{$\pm$ 0.07} & 0.96 \hl{$\pm$ 0.02} \\
& RATS & 0.913 $\pm$ 0.01 & 2.221 $\pm$ 0.51 & 0.91 \hl{$\pm$ 0.03} & 0.92 \hl{$\pm$ 0.02} \\
& RBET & 0.781 $\pm$ 0.10 & 3.628 $\pm$ 0.46 & 0.89 \hl{$\pm$ 0.05} & 0.70 \hl{$\pm$ 0.10} \\
\hline
\parbox[t]{2mm}{\multirow{6}{*}{\rotatebox[origin=c]{90}{CH}}}
& MedicDeepLabv3+ & 0.944 $\pm$ 0.04 & \hl{\textbf{2.064 $\pm$ 1.85}} & \textbf{0.94 \hl{$\pm$ 0.08}} & 0.96 \hl{$\pm$ 0.03} \\
& \hl{VoxResNet} & \hl{0.944 $\pm$ 0.04} & \hl{2.265 $\pm$ 1.86} & 0.93 \hl{$\pm$ 0.07} & 0.96 \hl{$\pm$ 0.02} \\
& \hl{HighRes3DNet} & \hl{0.942 $\pm$ 0.04} & \hl{2.205 $\pm$ 1.86} & 0.93 \hl{$\pm$ 0.07} & 0.96 \hl{$\pm$ 0.03} \\
& \hl{V-Net} & \hl{0.940 $\pm$ 0.04} & \hl{2.218 $\pm$ 1.86} & 0.93 \hl{$\pm$ 0.07} & 0.96 \hl{$\pm$ 0.03} \\
& UNet & 0.941 $\pm$ 0.05 & \hl{3.689 $\pm$ 1.64} & 0.92 \hl{$\pm$ 0.07} & \textbf{0.97 \hl{$\pm$ 0.02}} \\
& DeepLabv3+ & 0.921 $\pm$ 0.04 & \hl{2.411 $\pm$ 1.80} & 0.91 \hl{$\pm$ 0.07} & 0.94 \hl{$\pm$ 0.03} \\
\noalign{\smallskip}\hline
\end{tabular}
\end{center}
\end{table*}

\begin{figure*}[t]
\centering
\includegraphics[width=\textwidth]{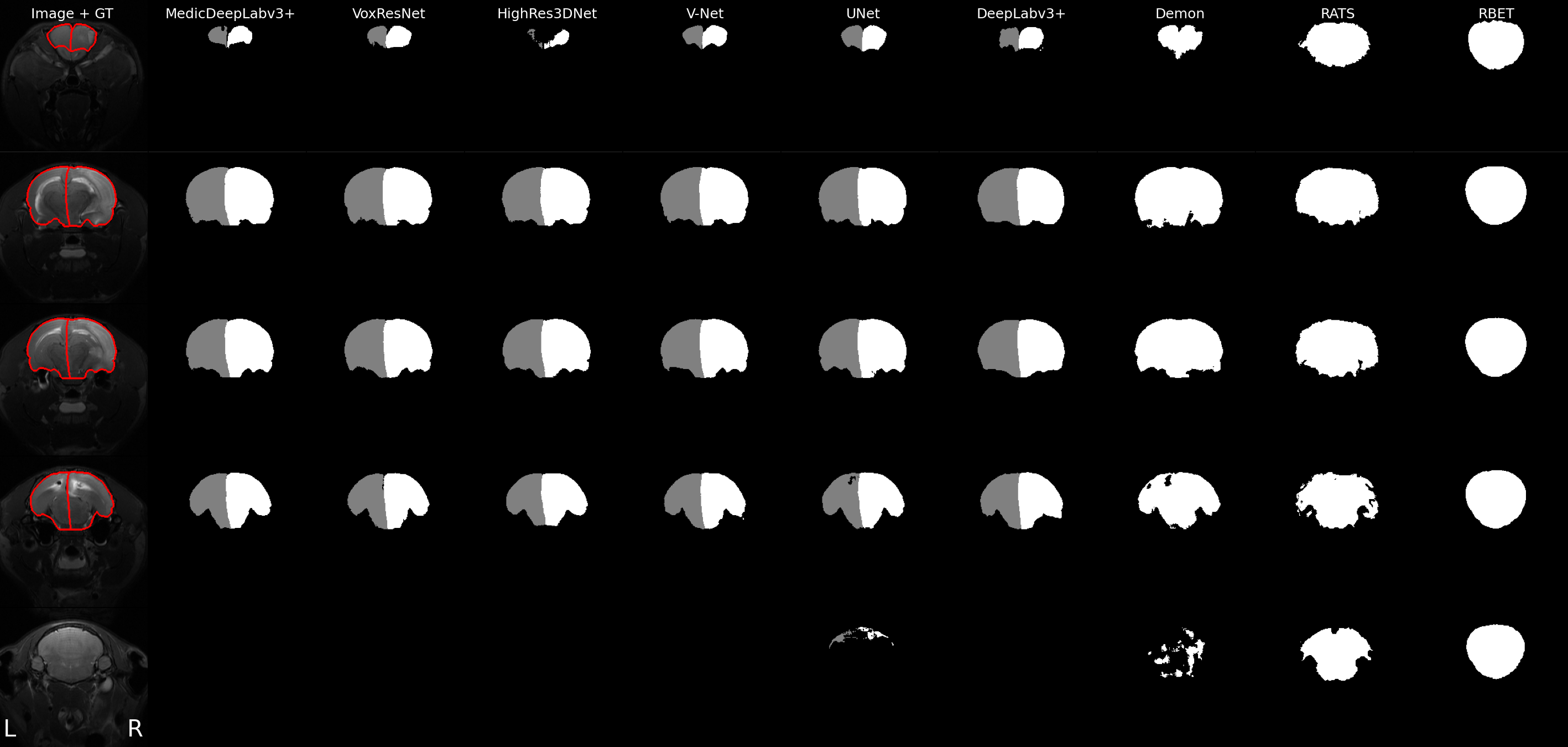}
\caption{T2-weighted image, ground truth and automatic segmentations of the rat with the most imbalanced hemispheric volumes.}
\label{fig:segmentations}
\end{figure*}

\subsection{Hemispheric ratio experiment}
The computed Cohen's d shows that, in terms of magnitude, all methods produced hemispheric ratio distributions not too different from the ground truth (Table \ref{table:pvalues}).
\hl{Among these methods, MedicDeepLabv3+ and V-Net provided the smallest effect size and the most zero-centered confidence interval, with MedicDeepLabv3+'s confidence interval being narrower than V-Net's.}
DeepLabv3+'s confidence interval was the largest and contained zero whereas UNet's confidence interval was the narrowest---slightly narrower than MedicDeepLabv3+'s---and did not contain zero.

\begin{table}[!t]
\caption{Cohen's d that measured the effect size and its confidence intervals.}
\label{table:pvalues}
\centering
\begin{tabular}{l c c}
\hline
Approach & Cohen's d & Confidence Interval  \\
\hline
MedicDeepLabv3+ & 0.008 & [-0.013, 0.035] \\
\hl{VoxResNet} & \hl{-0.042} & \hl{[-0.060, -0.025]} \\
\hl{HighRes3DNet} & \hl{-0.102} & \hl{[-0.125, -0.080]} \\
\hl{V-Net} & \hl{0.003} & \hl{[-0.042, 0.022]} \\
UNet & -0.038 & [-0.054, -0.021] \\
DeepLabv3+ & 0.050 & [-0.008, 0.099] \\
\hline
\end{tabular}
\end{table}

\subsection{Limited resources}
Table \ref{table:capacitycomparison} lists the characteristics, computational requirements, and performance of different versions of MedicDeepLabv3+ on the contralateral hemisphere segmentation (performance on the brain can be found in Online Resource 2).
Reducing the number of parameters by decreasing the number of initial filters reduced notably the required GPU memory and training time while it barely affected MedicDeepLabv3+'s performance.
For instance, reducing the number of parameters by 93.5\% (from 79.1M to 5.1M) decreased the required GPU memory and training time by 72\% while it decreased the Dice coefficient in the contralateral hemisphere by only 1\%.

Tables \ref{table:groupresults_brain} and \ref{table:groupresults_contra} show the performance of MedicDeepLabv3+ optimized and evaluated on each cohort and acquisition time-point separately.
In other words, for each cohort and acquisition time-point, the training set was comprised by only three images and test set size (Tables \ref{table:groupresults_brain} and \ref{table:groupresults_contra}, ``Volumes" column) varied across the 17 groups.
MedicDeepLabv3+, on average, performed slightly worse than in our first experiment that utilized 17 times more annotated data.
Performance measures across these groups varied notably: in the contralateral hemisphere segmentations (Table \ref{table:groupresults_contra}) Dice coefficients ranged from 0.876 to 0.951, HD from \hl{1.200 to 3.745}, precision from 0.871 to 0.962, and recall from 0.859 to 0.967.
Additionally, in agreement with our previous experiment, performance on the contralateral hemisphere was slightly lower than on the brain.

\begin{table*}
\begin{center}
\caption{Comparison between multiple versions of MedicDeepLabv3+ with different capacity. Columns: proportion of kernel filters with respect to the default configuration, trainable ConvNet parameters (in millions), optimization time for 300 epochs in our workstation (see Section \ref{sec:implementation} for details) in hours, maximum GPU memory required during training and evaluation, Dice and HD in the contralateral hemisphere \hl{(mean $\pm$ std)}. Bold: default configuration, highest performance.}
\label{table:capacitycomparison}
\begin{tabular}{lllllll}
\hline\noalign{\smallskip}
Rate & Parameters & Time (h) & Mem. (train) & Mem. (eval) & Dice & HD \\
\noalign{\smallskip}\hline\noalign{\smallskip}
\textbf{1} & \textbf{79.1M} & \textbf{16.2} & \textbf{8857 MiB} &\textbf{2935 MiB} & \textbf{0.944 \hl{$\pm$ 0.04}} & \textbf{2.064 \hl{$\pm$ 1.85}} \\
0.875 & 60.7M & 14.4 & 7571 MiB & 2617 MiB & 0.941 \hl{$\pm$ 0.04} & \hl{2.103 $\pm$ 1.86} \\
0.750 & 44.7M & 12.1 & 6545 MiB & 2319 MiB & 0.941 \hl{$\pm$ 0.04} & \hl{2.118 $\pm$ 1.86} \\
0.625 & 31.1M & 10.3 & 5619 MiB & 2007 MiB & 0.941 \hl{$\pm$ 0.04} & \hl{2.099 $\pm$ 1.85} \\
0.500 & 20.0M & 7.8 & 4577 MiB & 1717 MiB & 0.939 \hl{$\pm$ 0.05} & \hl{2.099 $\pm$ 1.83} \\
0.375 & 11.3M & 6.2 & 3531 MiB & 1421 MiB & 0.937 \hl{$\pm$ 0.05} & \hl{2.138 $\pm$ 1.85} \\
0.250 & 5.1M & 4.5 & 2503 MiB & 1121 MiB & 0.933 \hl{$\pm$ 0.05} & \hl{2.078 $\pm$ 1.82} \\
\noalign{\smallskip}\hline
\end{tabular}
\end{center}
\end{table*}

\begin{table*}
\begin{center}
\caption{Dice, Hausdorff distance (HD), precision, and recall on the brain masks derived with MedicDeepLabv3+ in each cohort and time-point (TP) separately. Volumes indicate the number of volumes in the test set. \hl{Mean $\pm$ std.}}
\label{table:groupresults_brain}
\begin{tabular}{lllllll}
\hline\noalign{\smallskip}
Cohort & TP & Volumes & Dice & HD & Prec & Recall \\
\noalign{\smallskip}\hline\noalign{\smallskip}
1 & 2h & 8 & 0.916 $\pm$ 0.03 & \hl{2.018 $\pm$ 0.47} & 0.929 \hl{$\pm$ 0.08} & 0.913 \hl{$\pm$ 0.08} \\ 
1 & 24h & 8 & 0.912 $\pm$ 0.10 & \hl{2.581 $\pm$ 1.82} & 0.894 \hl{$\pm$ 0.18} & 0.952 \hl{$\pm$ 0.02} \\ 
2 & 24h & 13 & 0.915 $\pm$ 0.02 & \hl{3.745 $\pm$ 1.12} & 0.929 \hl{$\pm$ 0.07} & 0.909 \hl{$\pm$ 0.05} \\ 
3 & D35 & 16 & 0.949 $\pm$ 0.02 &\hl{2.142 $\pm$ 0.44} & 0.956 \hl{$\pm$ 0.03} & 0.942 \hl{$\pm$ 0.02} \\ 
4 & 24h & 40 & 0.957 $\pm$ 0.01 & \hl{1.981 $\pm$ 0.87} & 0.985 \hl{$\pm$ 0.01} & 0.930 \hl{$\pm$ 0.02} \\ 
5 & 24h & 23 & 0.947 $\pm$ 0.01 & \hl{1.557 $\pm$ 0.43} & 0.935 \hl{$\pm$ 0.03} & 0.960 \hl{$\pm$ 0.02} \\ 
6 & D3 & 60 & 0.973 $\pm$ 0.01 & \hl{1.200 $\pm$ 0.32} & 0.968 \hl{$\pm$ 0.02} & 0.978 \hl{$\pm$ 0.01} \\ 
6 & D28 & 58 & 0.946 $\pm$ 0.02 & \hl{1.688 $\pm$ 0.60} & 0.937 \hl{$\pm$ 0.04} & 0.957 \hl{$\pm$ 0.02} \\ 
7 & D3 & 35 & 0.956 $\pm$ 0.01 & \hl{1.723 $\pm$ 0.92} & 0.968 \hl{$\pm$ 0.02} & 0.945 \hl{$\pm$ 0.03} \\ 
7 & D21 & 35 & 0.950 $\pm$ 0.01 & \hl{1.548 $\pm$ 0.75} & 0.937 \hl{$\pm$ 0.02} & 0.964 \hl{$\pm$ 0.01} \\ 
8 & 24h & 29 & 0.956 $\pm$ 0.01 & \hl{1.742 $\pm$ 0.92} & 0.956 \hl{$\pm$ 0.03} & 0.956 \hl{$\pm$ 0.02} \\ 
8 & D3 & 26 & 0.953 $\pm$ 0.01 & \hl{1.599 $\pm$ 0.35} & 0.949 \hl{$\pm$ 0.02} & 0.957 \hl{$\pm$ 0.01} \\ 
8 & D14 & 26 & 0.948 $\pm$ 0.01 & \hl{1.678 $\pm$ 0.43} & 0.929 \hl{$\pm$ 0.03} & 0.967 \hl{$\pm$ 0.01} \\ 
8 & D28 & 23 & 0.943 $\pm$ 0.02 & \hl{1.713 $\pm$ 0.36} & 0.940 \hl{$\pm$ 0.03} & 0.946 \hl{$\pm$ 0.03} \\ 
9 & 24h & 77 & 0.951 $\pm$ 0.03 & \hl{1.838 $\pm$ 1.02} & 0.970 \hl{$\pm$ 0.04} & 0.935 \hl{$\pm$ 0.04} \\ 
10 & D7 & 36 & 0.919 $\pm$ 0.05 & \hl{2.226 $\pm$ 0.81} & 0.880 \hl{$\pm$ 0.10} & 0.970 \hl{$\pm$ 0.02} \\ 
11 & 24h & 28 & 0.937 $\pm$ 0.02 & \hl{2.696 $\pm$ 0.79} & 0.954 \hl{$\pm$ 0.04} & 0.923 \hl{$\pm$ 0.04} \\ 
\hline
\multicolumn{2}{l}{Average} & 541 & 0.948 $\pm$ 0.03 & \hl{1.833 $\pm$ 0.89} & 0.949 \hl{$\pm$ 0.05} & 0.951 \hl{$\pm$ 0.03} \\ 
\noalign{\smallskip}\hline
\end{tabular}
\end{center}
\end{table*}

\begin{table*}
\begin{center}
\caption{Dice, Hausdorff distance (HD), precision, and recall on the contralateral hemisphere masks derived with MedicDeepLabv3+ in each cohort and time-point (TP) separately. Volumes indicate the number of volumes in the test set. \hl{Mean $\pm$ std.}}
\label{table:groupresults_contra}
\begin{tabular}{lllllll}
\hline\noalign{\smallskip}
Cohort & TP & Volumes & Dice & HD & Prec & Recall \\
\noalign{\smallskip}\hline\noalign{\smallskip}
1 & 2h & 8 & 0.883 $\pm$ 0.03 & \hl{3.593 $\pm$ 0.83} & 0.871 \hl{$\pm$ 0.06} & 0.904 \hl{$\pm$ 0.07} \\  
1 & 24h & 8 & 0.886 $\pm$ 0.11 & \hl{3.181 $\pm$ 1.76} & 0.874 \hl{$\pm$ 0.18} & 0.915 \hl{$\pm$ 0.03} \\  
2 & 24h & 13 & 0.876 $\pm$ 0.03 & \hl{3.792 $\pm$ 1.60} & 0.902 \hl{$\pm$ 0.08} & 0.859 \hl{$\pm$ 0.05} \\  
3 & D35 & 16 & 0.927 $\pm$ 0.02 & \hl{1.977 $\pm$ 0.94} & 0.948 \hl{$\pm$ 0.03} & 0.907 \hl{$\pm$ 0.03} \\  
4 & 24h & 40 & 0.928 $\pm$ 0.02 & \hl{1.889 $\pm$ 0.76} & 0.962 \hl{$\pm$ 0.03} & 0.898 \hl{$\pm$ 0.03} \\  
5 & 24h & 23 & 0.899 $\pm$ 0.04 & \hl{1.630 $\pm$ 0.48} & 0.912 \hl{$\pm$ 0.04} & 0.888 \hl{$\pm$ 0.06} \\  
6 & D3 & 60 & 0.951 $\pm$ 0.02 & \hl{2.766 $\pm$ 2.27} & 0.936 \hl{$\pm$ 0.04} & 0.967 \hl{$\pm$ 0.01} \\  
6 & D28 & 58 & 0.935 $\pm$ 0.02 & \hl{1.387 $\pm$ 0.51} & 0.932 \hl{$\pm$ 0.04} & 0.939 \hl{$\pm$ 0.02} \\  
7 & D3 & 35 & 0.930 $\pm$ 0.02 & \hl{2.038 $\pm$ 1.40} & 0.959 \hl{$\pm$ 0.02} & 0.904 \hl{$\pm$ 0.04} \\  
7 & D21 & 35 & 0.939 $\pm$ 0.01 & \hl{1.487 $\pm$ 0.90} & 0.927 \hl{$\pm$ 0.03} & 0.952 \hl{$\pm$ 0.02} \\  
8 & 24h & 29 & 0.935 $\pm$ 0.02 & \hl{2.574 $\pm$ 2.32} & 0.936 \hl{$\pm$ 0.03} & 0.936 \hl{$\pm$ 0.03} \\  
8 & D3 & 26 & 0.903 $\pm$ 0.07 & \hl{5.874 $\pm$ 0.82} & 0.902 \hl{$\pm$ 0.03} & 0.912 \hl{$\pm$ 0.11} \\  
8 & D14 & 26 & 0.933 $\pm$ 0.02 & \hl{1.911 $\pm$ 1.34} & 0.901 \hl{$\pm$ 0.04} & 0.967 \hl{$\pm$ 0.01} \\  
8 & D28 & 23 & 0.932 $\pm$ 0.02 & \hl{1.830 $\pm$ 1.23} & 0.929 \hl{$\pm$ 0.03} & 0.935 \hl{$\pm$ 0.03} \\  
9 & 24h & 77 & 0.917 $\pm$ 0.03 & \hl{2.234 $\pm$ 1.67} & 0.926 \hl{$\pm$ 0.05} & 0.911 \hl{$\pm$ 0.04} \\  
10 & D7 & 36 & 0.908 $\pm$ 0.05 & \hl{3.195 $\pm$ 2.20} & 0.871 \hl{$\pm$ 0.10} & 0.958 \hl{$\pm$ 0.03} \\  
11 & 24h & 28 & 0.894 $\pm$ 0.03 & \hl{4.200 $\pm$ 2.27} & 0.900 \hl{$\pm$ 0.04} & 0.892 \hl{$\pm$ 0.04} \\  
\hline
\multicolumn{2}{l}{Average} & 541 & 0.923 $\pm$ 0.04 & \hl{2.480 $\pm$ 1.89} & 0.924 \hl{$\pm$ 0.05} & 0.926 \hl{$\pm$ 0.05} \\
\noalign{\smallskip}\hline
\end{tabular}
\end{center}
\end{table*}

\section{Discussion}
We presented MedicDeepLabv3+, the first method for hemisphere segmentation in \hl{rat} MR images with lesions.
We compared MedicDeepLabv3+ performance with state-of-the-art DeepLabv3+, UNet, \hl{HighRes3DNet, V-Net, VoxResNet,} and three brain extraction algorithms (Demon, RATS, and RBET) combining several preclinical neuroimaging studies to a large dataset of 723 rat MR volumes.

ConvNets performed markedly better and their training time was about 10 times shorter than RATS \citep{oguz2014rats} and RBET \citep{wood2013rbet}.
The superior performance of ConvNets was not surprising, \hl{as RATS and RBET were not designed to segment brains with widely varying intensity values, such as those found in brains with lesions}.
This outperformance of ConvNets over more traditional segmentation algorithms on rodent MRI aligns with recent research \citep{roy2018deep,liu2020automatic,de2021automated}.

MedicDeepLabv3+ yielded the highest Dice coefficients, precision and recall, and the lowest HD (Table \ref{table:segmentationcomparison}).
Particularly, the outperformance of MedicDeepLabv3+ over the baseline DeepLabv3+ \citep{chen2018encoder} indicates that our proposed modifications (i.e., the incorporation of spatial attention layers and additional skip-connections) led to improvements.
Similar improvements after incorporating attention layers, such as the proposed spatial attention layers, have also been reported in the literature \citep{oktay2018attention,wang2019automatic,tao2019improving,xu2020asymmetrical}.
In the brain midline area experiment, UNet achieved slightly higher Dice coefficients than \hl{the other 3D ConvNets}.
However, these Dice coefficients were computed only in the annotated slices, as finding the brain midline requires the manual annotations.
As we showed in Table \ref{table:segmentationcomparison}, Figure \ref{fig:segmentations}, and the 17 Figures in Online Resource 1, 2D ConvNets, including UNet, produced misclassifications in the cerebellum and the olfactory bulb that were not annotated, leading to notably higher HD.
Therefore, the small difference between UNet and \hl{the 3D ConvNets} (Fig. \ref{fig:brainmidline}) comes at the expense of those misclassifications that were disregarded during the evaluation.
\hl{The differences among HighRes3DNet, V-Net, VoxResNet and MedicDeepLabv3+ were also very small}.
In contrast, the difference between MedicDeepLabv3+ and the baseline DeepLabv3+ was three times larger than between MedicDeepLabv3+ and UNet.

Our benchmark (Table \ref{table:segmentationcomparison}) provides a valuable insight into whether 2D ConvNets produce better segmentations than 3D ConvNets on highly anisotropic data.
In recent literature, 2D ConvNets appeared to be better \citep{jang2017automatic,isensee2017automatic,baumgartner2017exploration}, including in rodent images similar to our dataset \citep{de2021automated}.
2D ConvNets outperformance may arise because contiguous slices can differ significantly in anisotropic data, thus, three-dimensional information might be unnecessary, and slice appearance might suffice to segment the regions of interest.
Our data and, particularly, our manual annotations, were specially challenging since our regions of interest had similar intensity values to the cerebellum and olfactory bulb that were not annotated.
Therefore, three-dimensional information can be critical to learn the location in the rostro-caudal axis of certain areas to avoid them.
Indeed, our results support this intuition.
Although Dice coefficient, precision and recall varied across architectures (Table \ref{table:segmentationcomparison}), HD was consistently lower with 3D ConvNets.
In other words, 2D ConvNets produced more critical misclassifications.
Thus, our data showcased a scenario in which, despite the anisotropy, 3D ConvNets were superior to 2D ConvNets, showing that the architectural choices need to consider more specific information and not just whether the data is anisotropic.

We measured the discrepancy magnitude between the hemispheric ratio distributions from the segmentations and from the ground truth (Table \ref{table:pvalues}), and \hl{V-Net and} our MedicDeepLabv3+ yielded the smallest effect size, indicating that the hemispheric ratios of their corresponding segmentations were more similar to the ground truth than the other ConvNets.
We want to emphasize the importance of accurate hemispheric ratios as they are biomarkers for predicting acute stroke \citep{swanson1990semiautomated,gerriets2004noninvasive}.
Both V-Net and MedicDeepLabv3+'s confidence intervals were zero-centered, and between these two, MedicDeepLabv3+'s was one third smaller than V-Net's.
\hl{The effect size of VoxResNet and HighRes3DNet (the second and third best performing ConvNets after MedicDeepLabv3+) was much higher, and their confidence intervals did not include zero, which indicates that their hemispheric ratios were biased, being considerably larger than the ground truth.
UNet's and DeepLabv3+'s effect size were also high, and DeepLabv3+'s confidence interval was the largest across all compared ConvNets.}

ConvNets, and especially our MedicDeepLabv3+, produced segmentations more similar to the ground truth than the other methods (Table \ref{table:segmentationcomparison}).
Since these ConvNets were high capacity---requiring large GPU memory---and they were optimized with several images, their outperformance is in line with recent research \citep{tan2019efficientnet}.
However, annotated data are often scarce, and large GPU memory to optimize ConvNets is not necessarily available.
Motivated by these constraints, we showed in two separate experiments that  MedicDeepLabv3+ performed remarkably well with few annotated data and very limited GPU memory (see Tables \ref{table:capacitycomparison}, \ref{table:groupresults_brain}, and \ref{table:groupresults_contra}).
In other words, our method can handle different scenarios without excessively sacrificing performance, which showcases MedicDeepLabv3+ generalization capabilities.

MedicDeepLabv3+ is publicly available, and it can be easily incorporated into existing pipelines, reducing human workload and accelerating rodent neuroimaging analyses.
Furthermore, MedicDeepLabv3+ is fast, requires no preprocessing and postprocessing, and it can be optimized on MR images with different contrast, voxel resolution, field of view, lesion appearance, and limited GPU memory and annotated data.
As hemisphere segmentation masks can be utilized in diverse studies, our work is relevant for multiple applications involving brain lesions in \hl{rat} images.

\section*{Acknowledgements}
The work of J.M. Valverde was funded from the European Union's Horizon 2020 Framework Programme (Marie Skłodowska Curie grant agreement \#740264 (GENOMMED)).
This work has also been supported by the grant \#316258 from Academy of Finland (J. Tohka) and grant S21770 from the European Social Fund (R. De Feo).
Part of the computational analysis was run on the servers provided by Bioinformatics Center, University of Eastern Finland, Finland.

\bibliography{mybibfile}

\newpage
\onecolumn

{\Large Appendix}

\setcounter{section}{0}
\renewcommand\thesection{\Alph{section}}

\section{Supplementary Figures}

\begin{figure*}[b!]
\centering
\includegraphics[width=\textwidth]{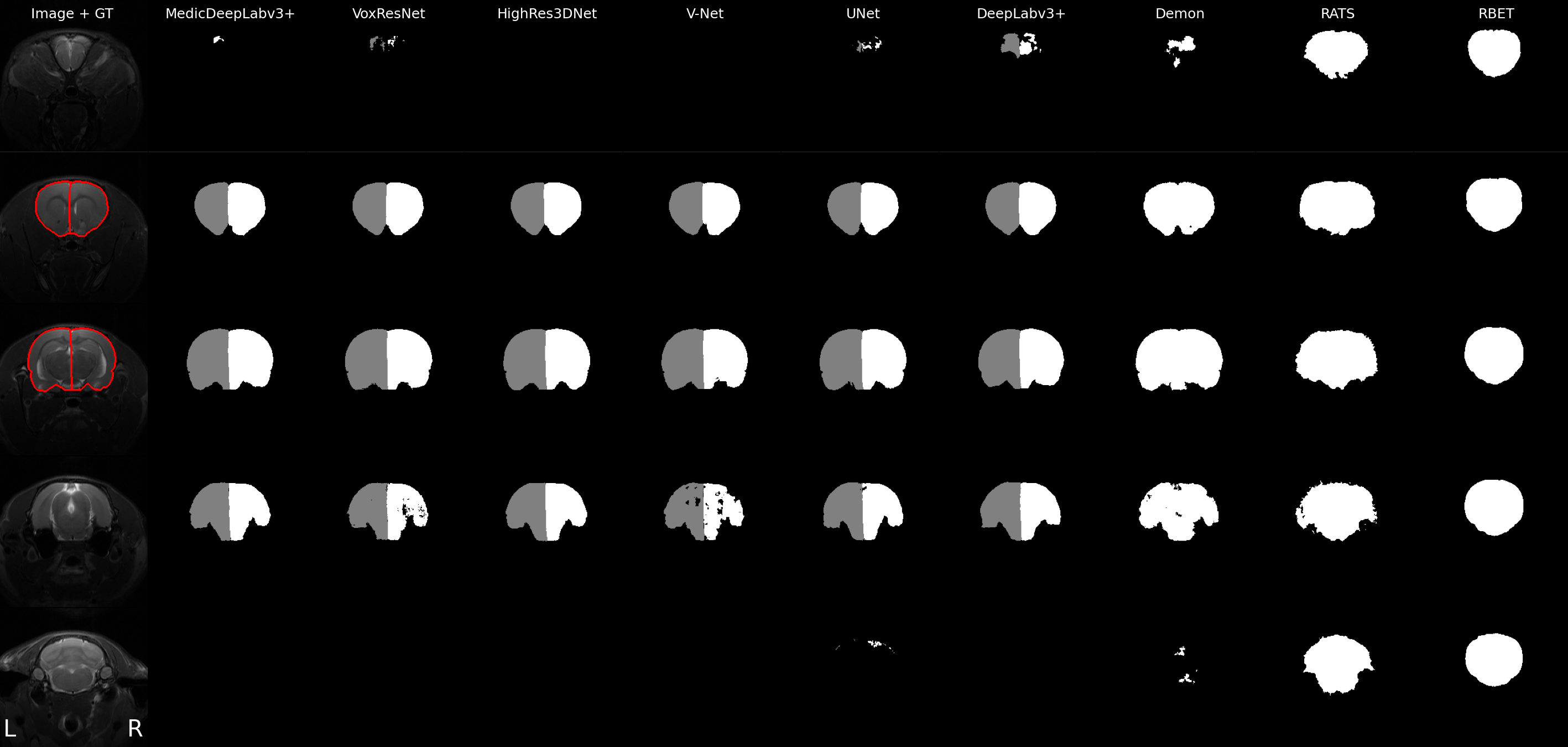}
\caption{T2-weighted image, ground truth and automatic segmentations of a rat from Study 1, 2h.}
\label{fig:segmentations}
\end{figure*}

\begin{figure*}[bt!]
\centering
\includegraphics[width=\textwidth]{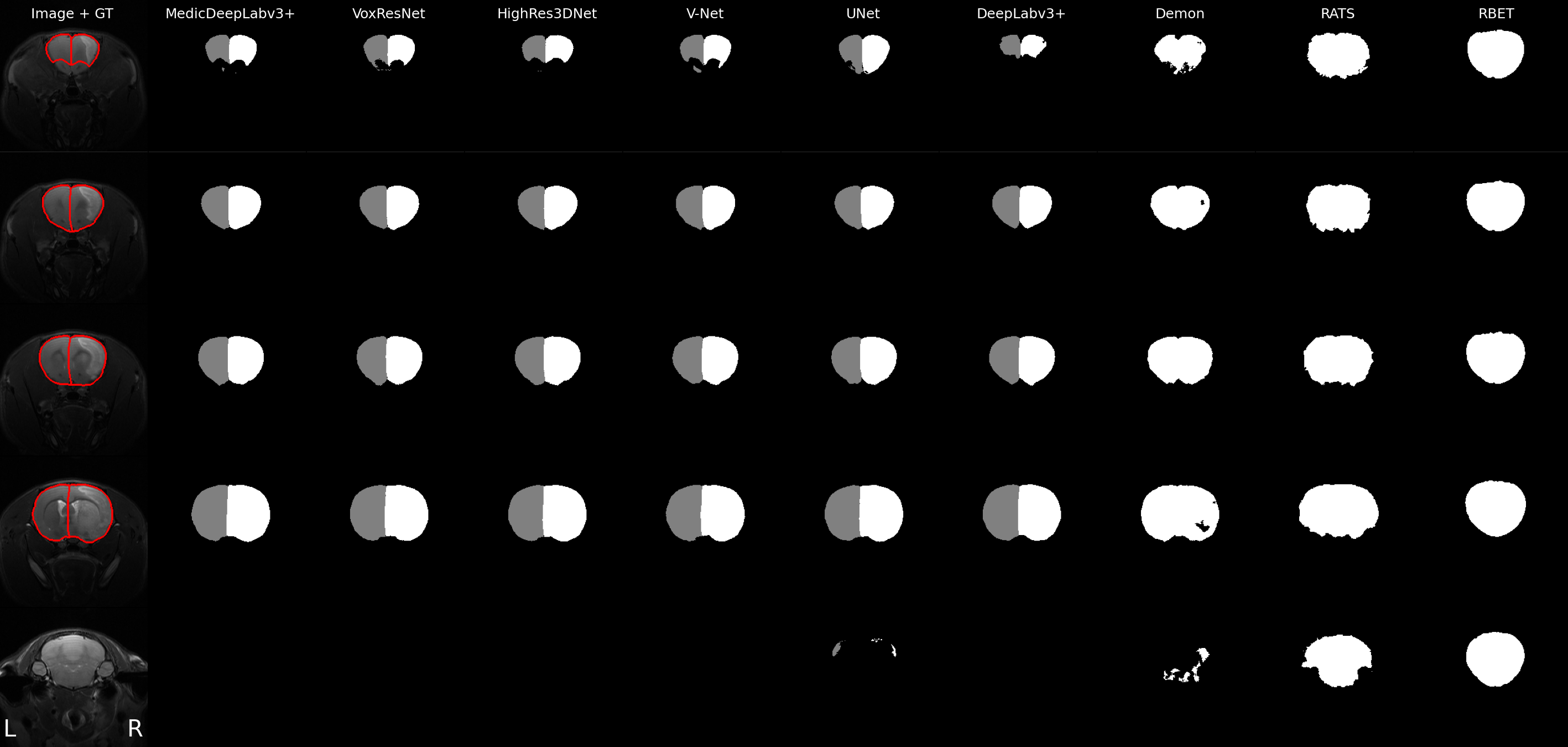}
\caption{T2-weighted image, ground truth and automatic segmentations of a rat from Study 1, 24h.}
\label{fig:segmentations}
\end{figure*}

\begin{figure*}[bt!]
\centering
\includegraphics[width=\textwidth]{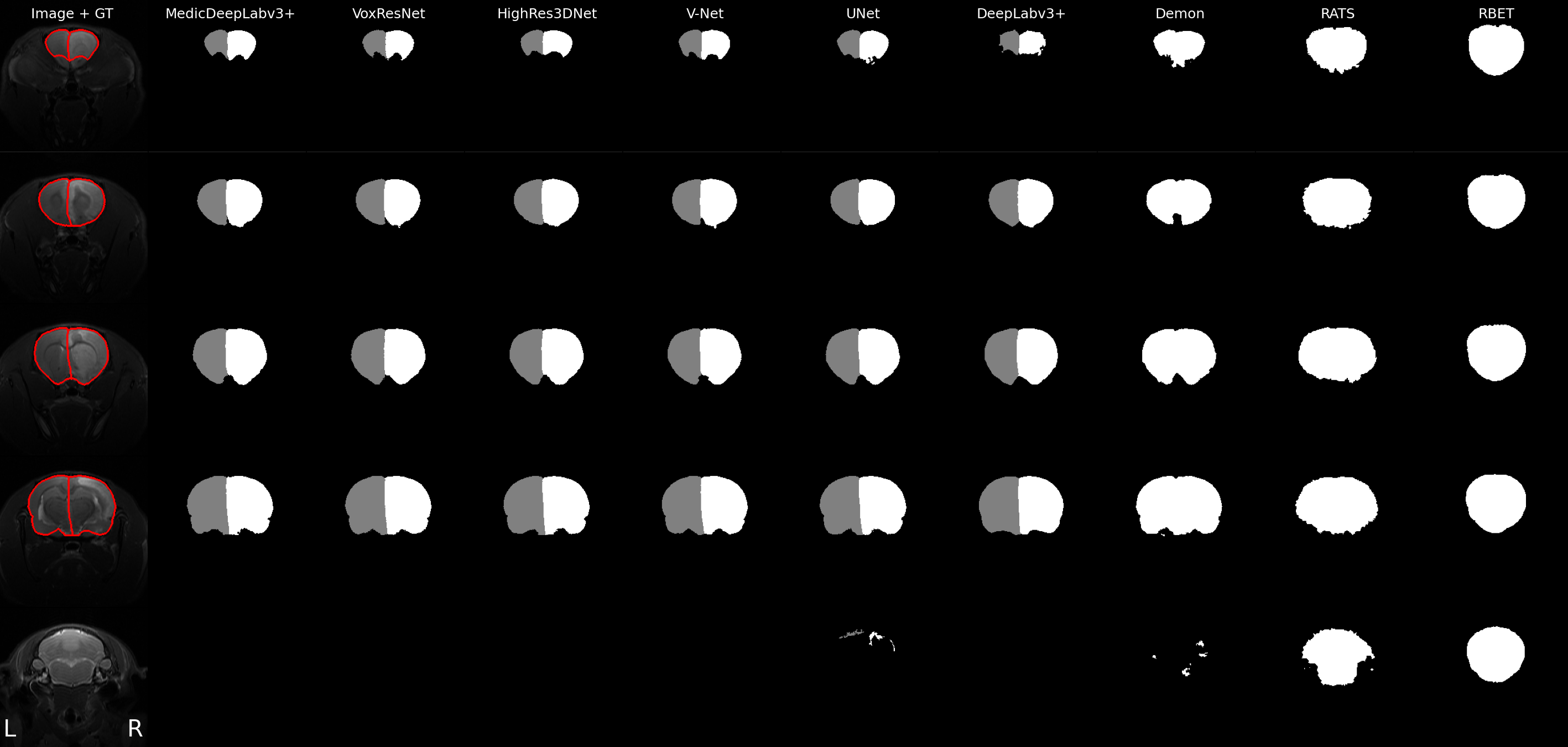}
\caption{T2-weighted image, ground truth and automatic segmentations of a rat from Study 2, 24h.}
\label{fig:segmentations}
\end{figure*}

\begin{figure*}[bt!]
\centering
\includegraphics[width=\textwidth]{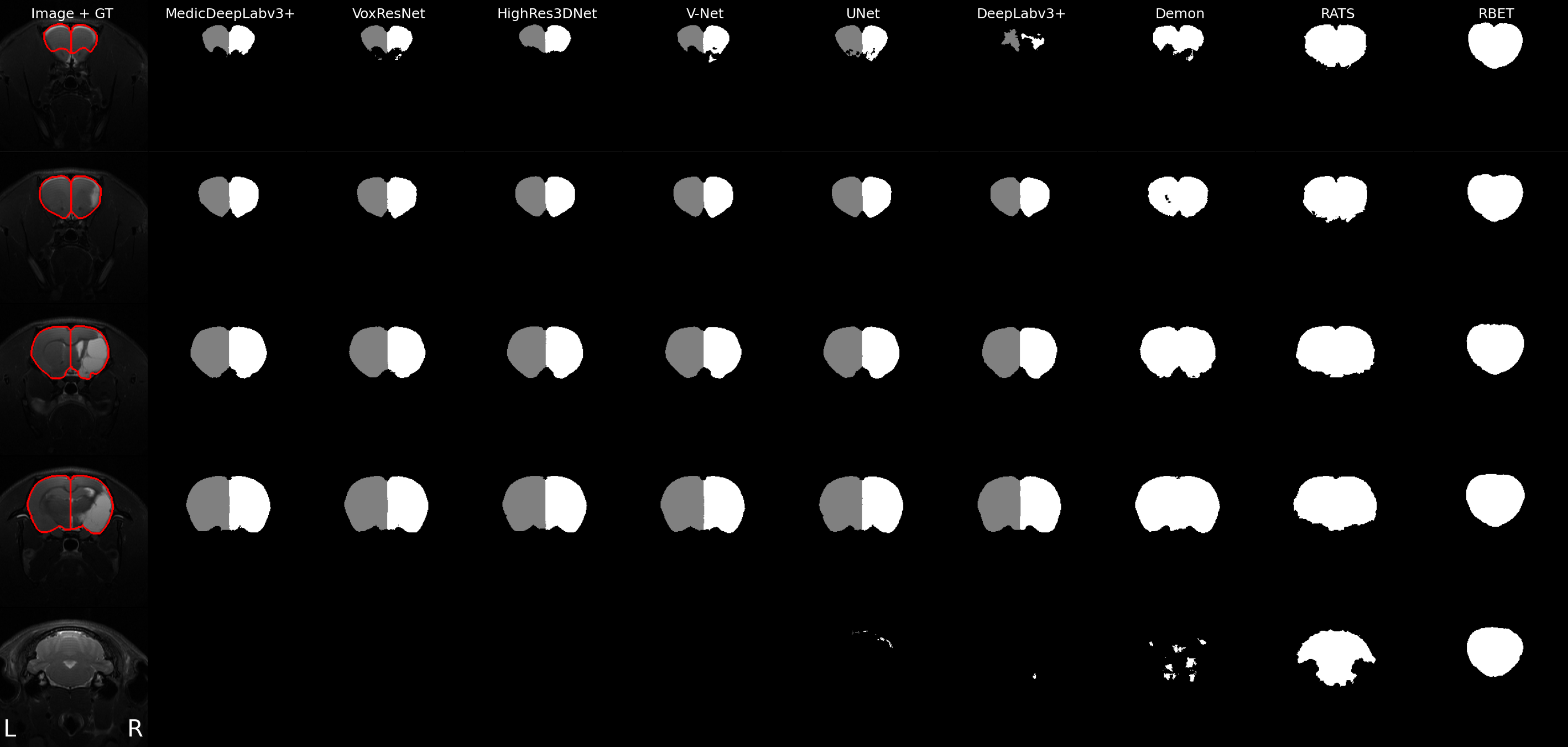}
\caption{T2-weighted image, ground truth and automatic segmentations of a rat from Study 3, D35.}
\label{fig:segmentations}
\end{figure*}

\begin{figure*}[bt!]
\centering
\includegraphics[width=\textwidth]{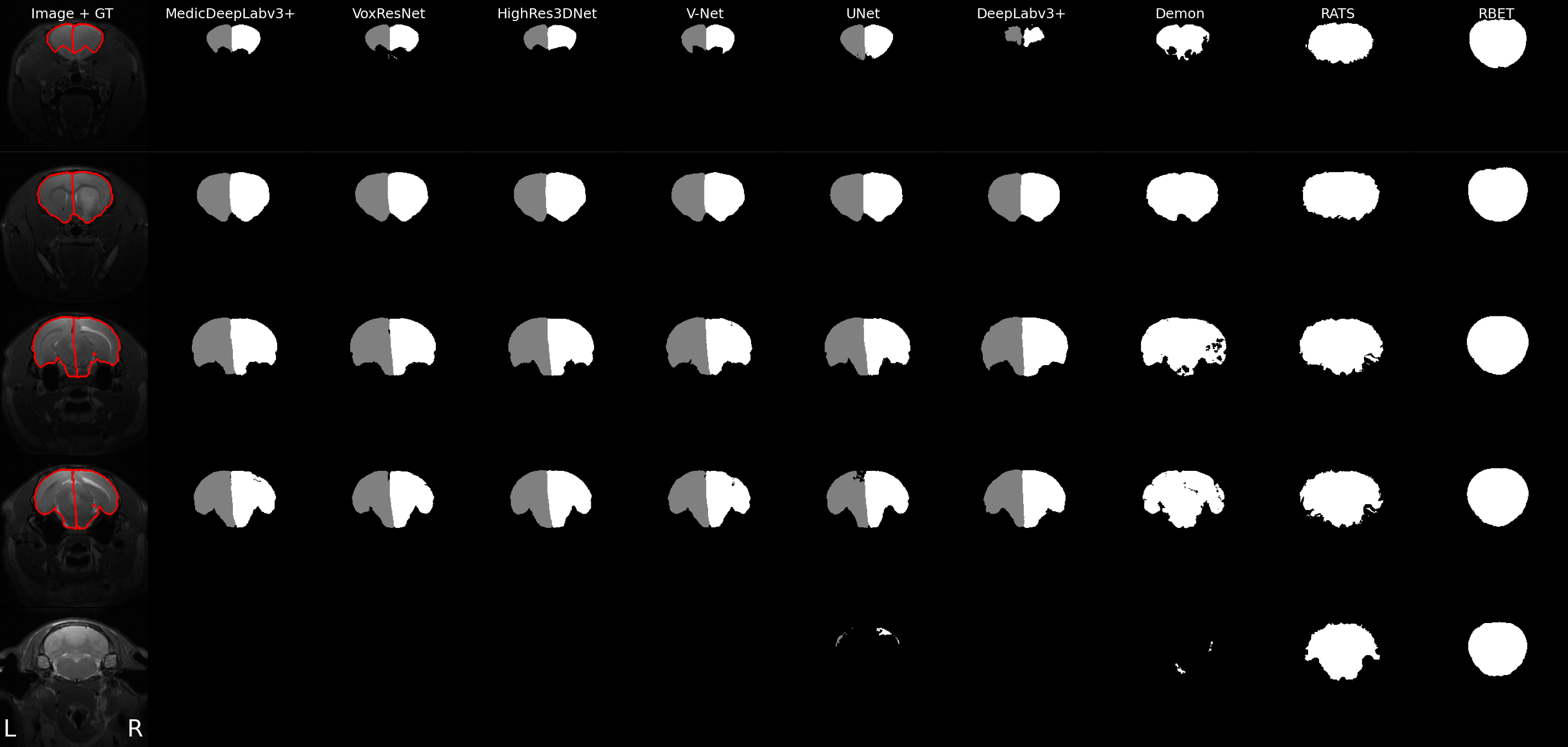}
\caption{T2-weighted image, ground truth and automatic segmentations of a rat from Study 4, 24h.}
\label{fig:segmentations}
\end{figure*}

\begin{figure*}[bt!]
\centering
\includegraphics[width=\textwidth]{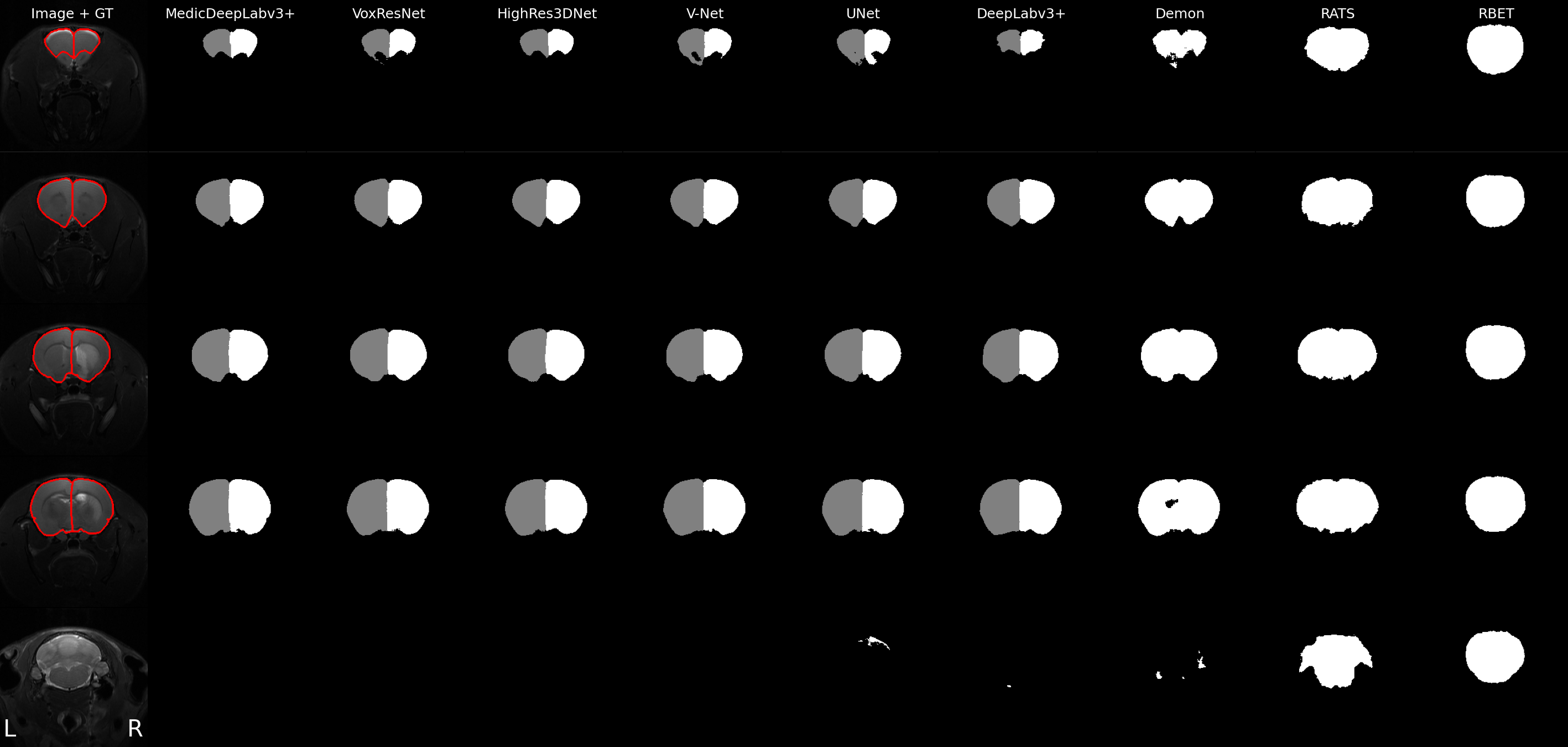}
\caption{T2-weighted image, ground truth and automatic segmentations of a rat from Study 5, 24h.}
\label{fig:segmentations}
\end{figure*}

\begin{figure*}[bt!]
\centering
\includegraphics[width=\textwidth]{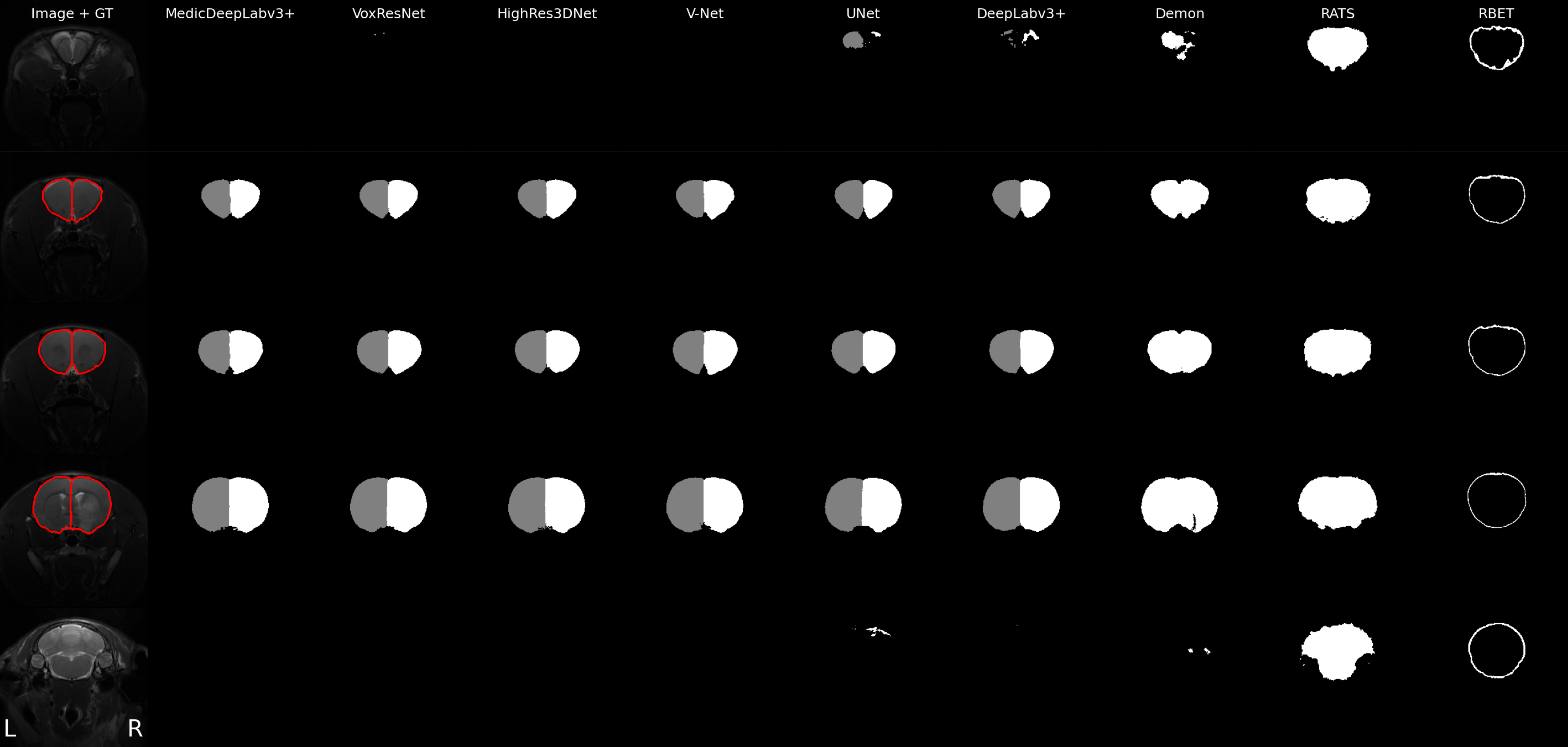}
\caption{T2-weighted image, ground truth and automatic segmentations of a rat from Study 6, D3.}
\label{fig:segmentations}
\end{figure*}

\begin{figure*}[bt!]
\centering
\includegraphics[width=\textwidth]{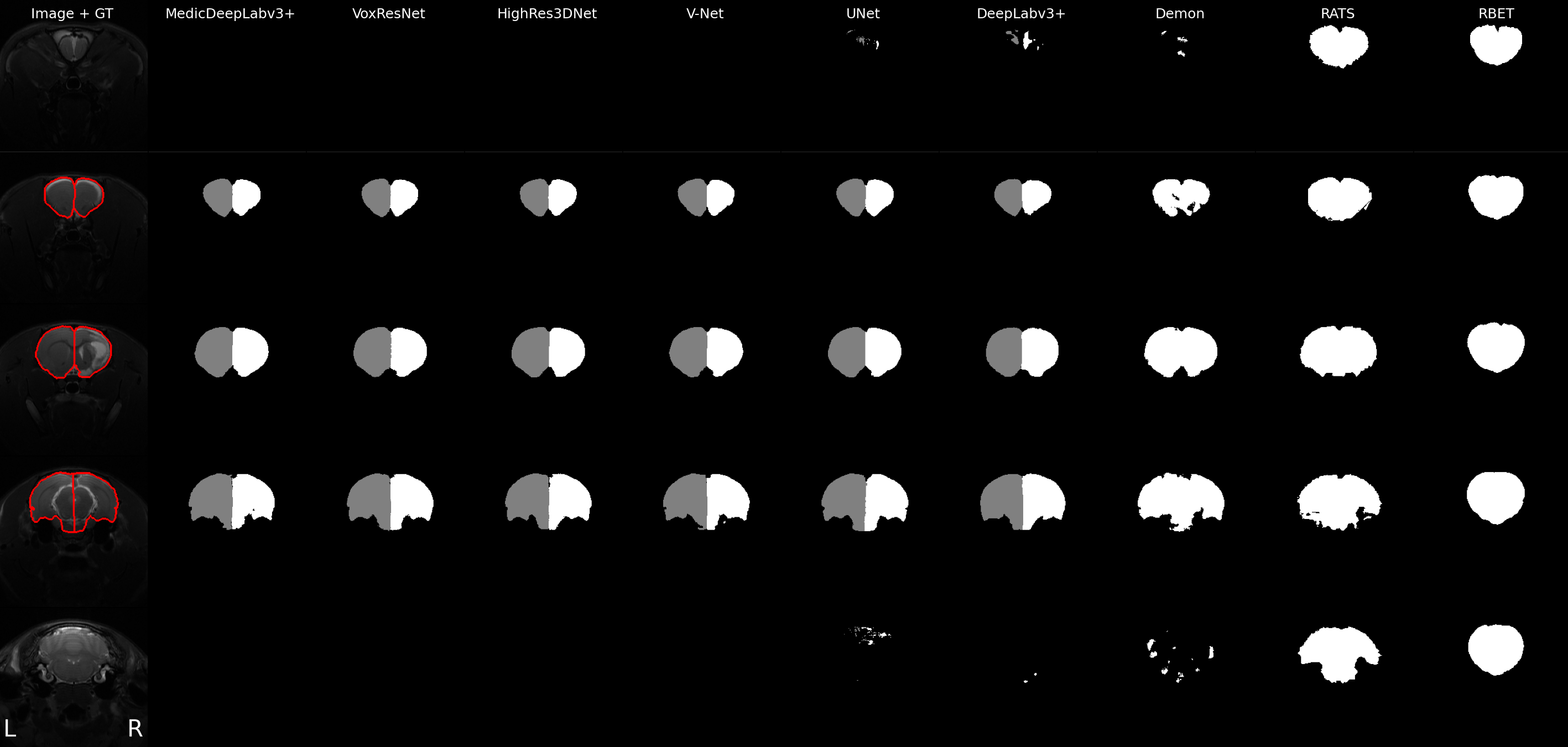}
\caption{T2-weighted image, ground truth and automatic segmentations of a rat from Study 6, D28.}
\label{fig:segmentations}
\end{figure*}

\begin{figure*}[bt!]
\centering
\includegraphics[width=\textwidth]{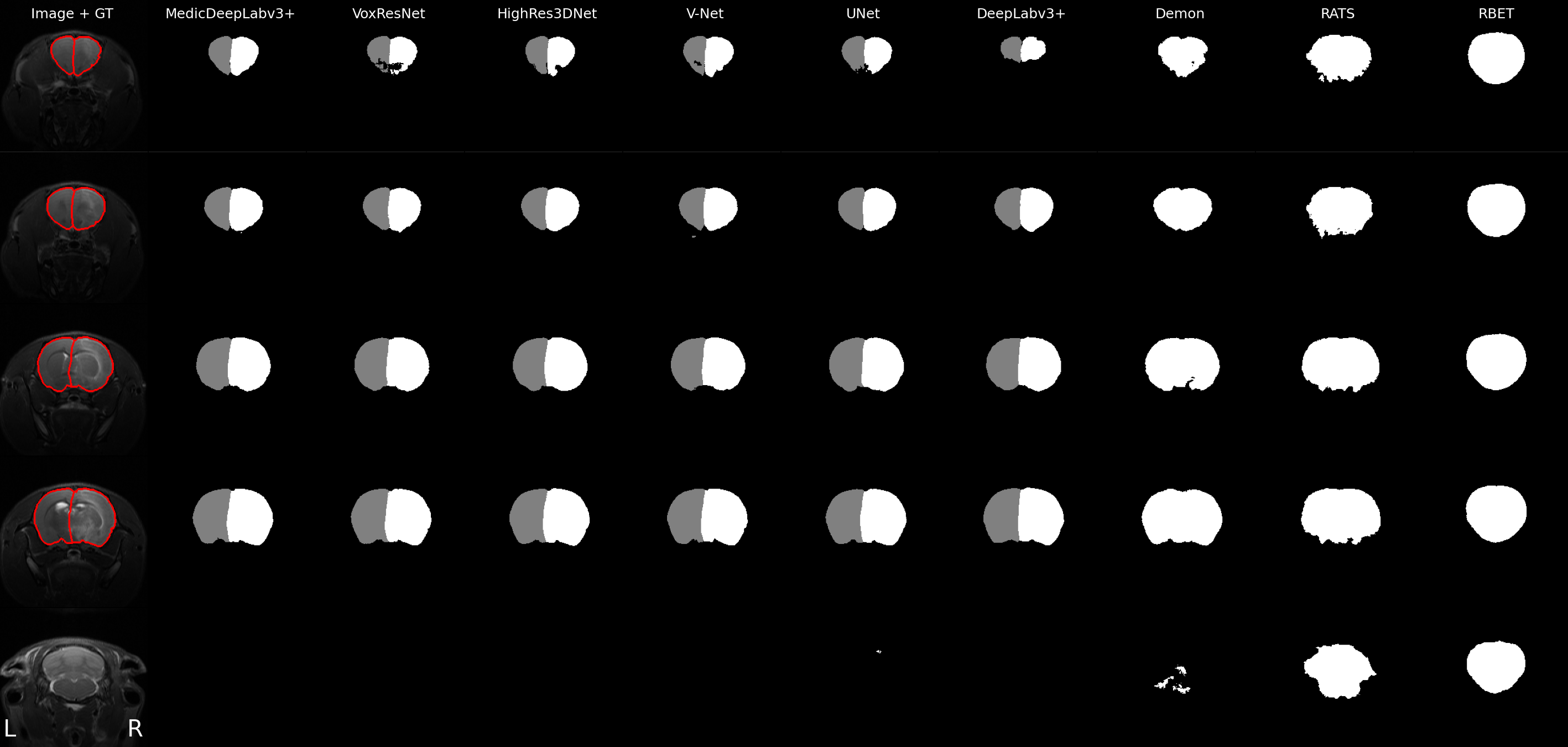}
\caption{T2-weighted image, ground truth and automatic segmentations of a rat from Study 7, D3.}
\label{fig:segmentations}
\end{figure*}

\begin{figure*}[bt!]
\centering
\includegraphics[width=\textwidth]{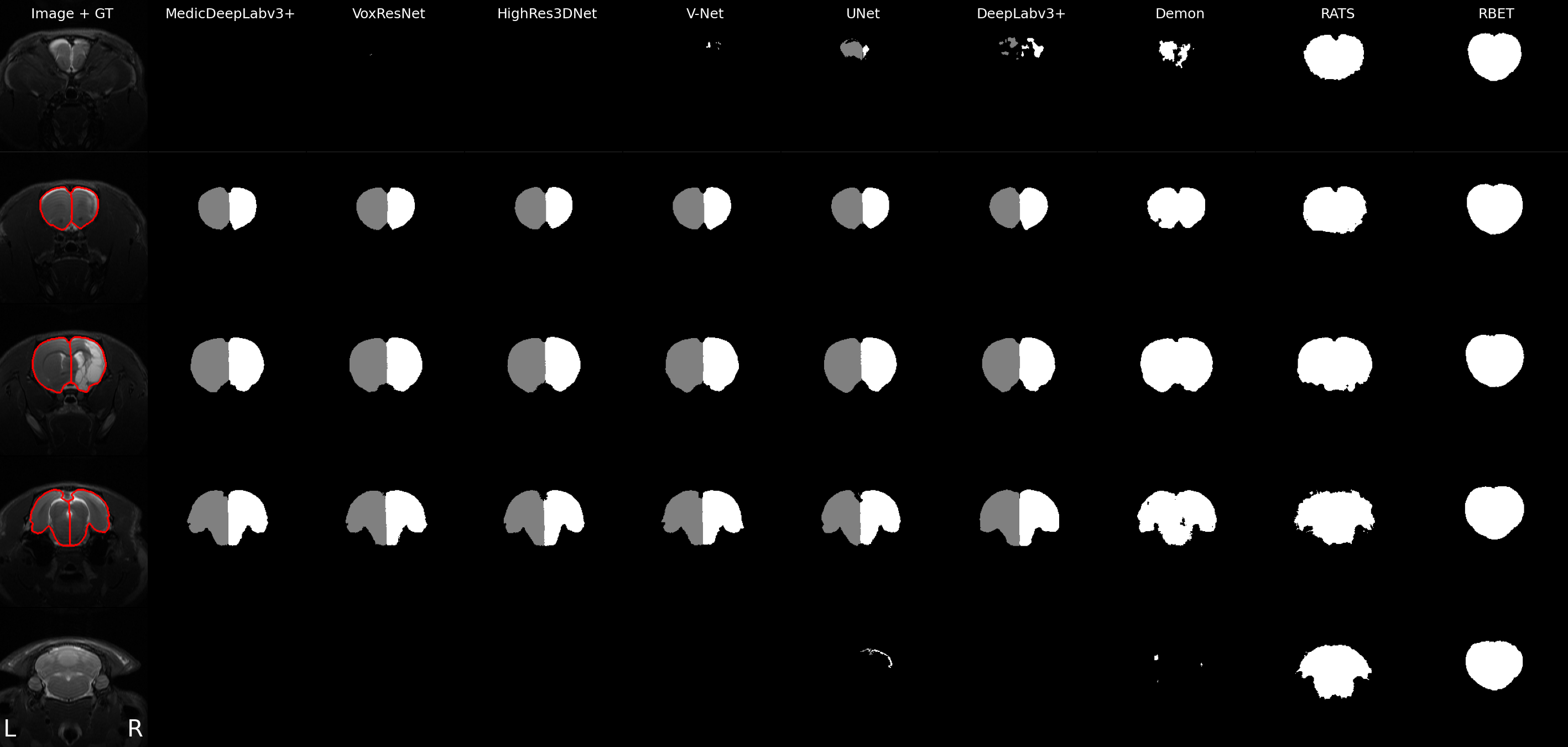}
\caption{T2-weighted image, ground truth and automatic segmentations of a rat from Study 7, D21.}
\label{fig:segmentations}
\end{figure*}

\begin{figure*}[bt!]
\centering
\includegraphics[width=\textwidth]{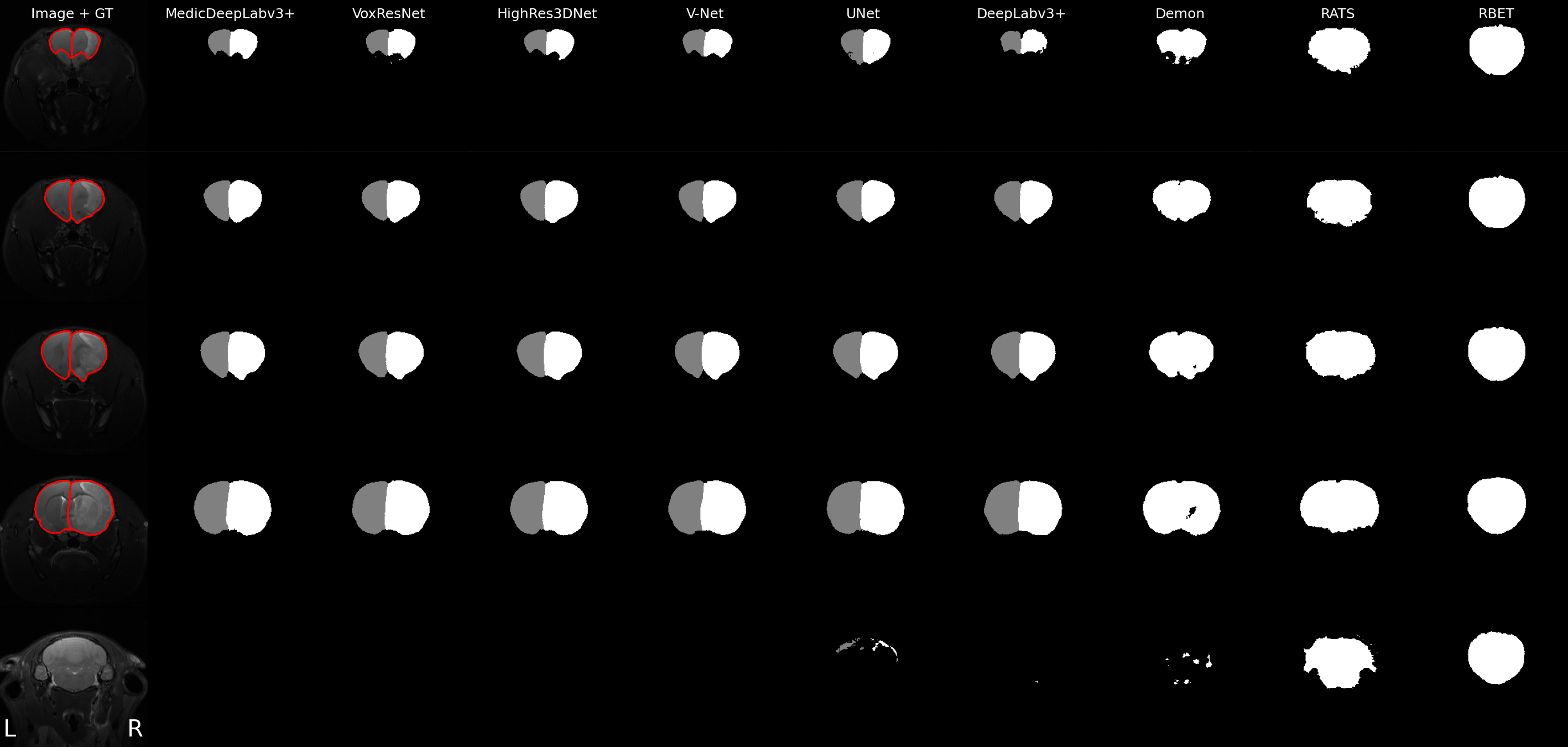}
\caption{T2-weighted image, ground truth and automatic segmentations of a rat from Study 8, 24h.}
\label{fig:segmentations}
\end{figure*}

\begin{figure*}[bt!]
\centering
\includegraphics[width=\textwidth]{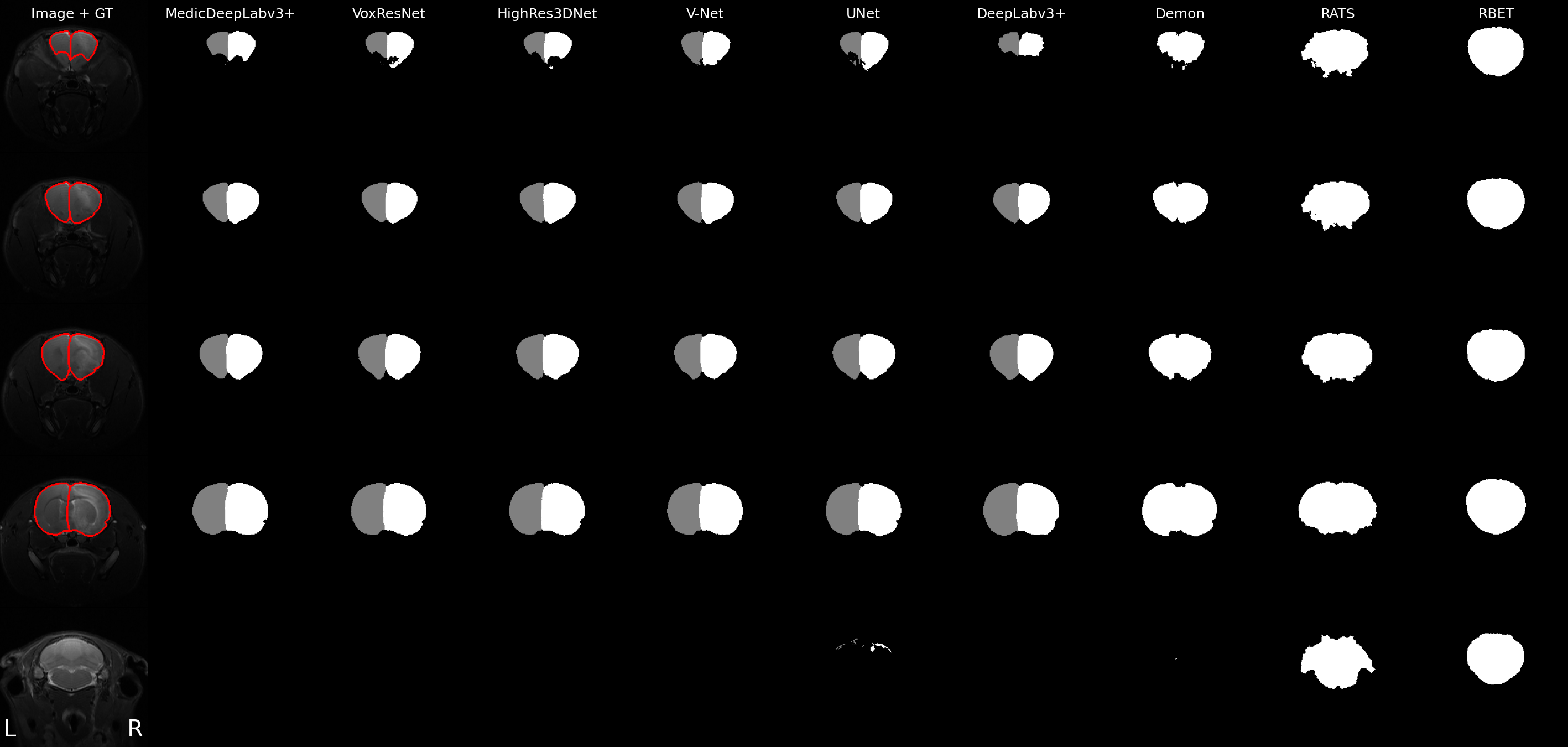}
\caption{T2-weighted image, ground truth and automatic segmentations of a rat from Study 8, D3.}
\label{fig:segmentations}
\end{figure*}

\begin{figure*}[bt!]
\centering
\includegraphics[width=\textwidth]{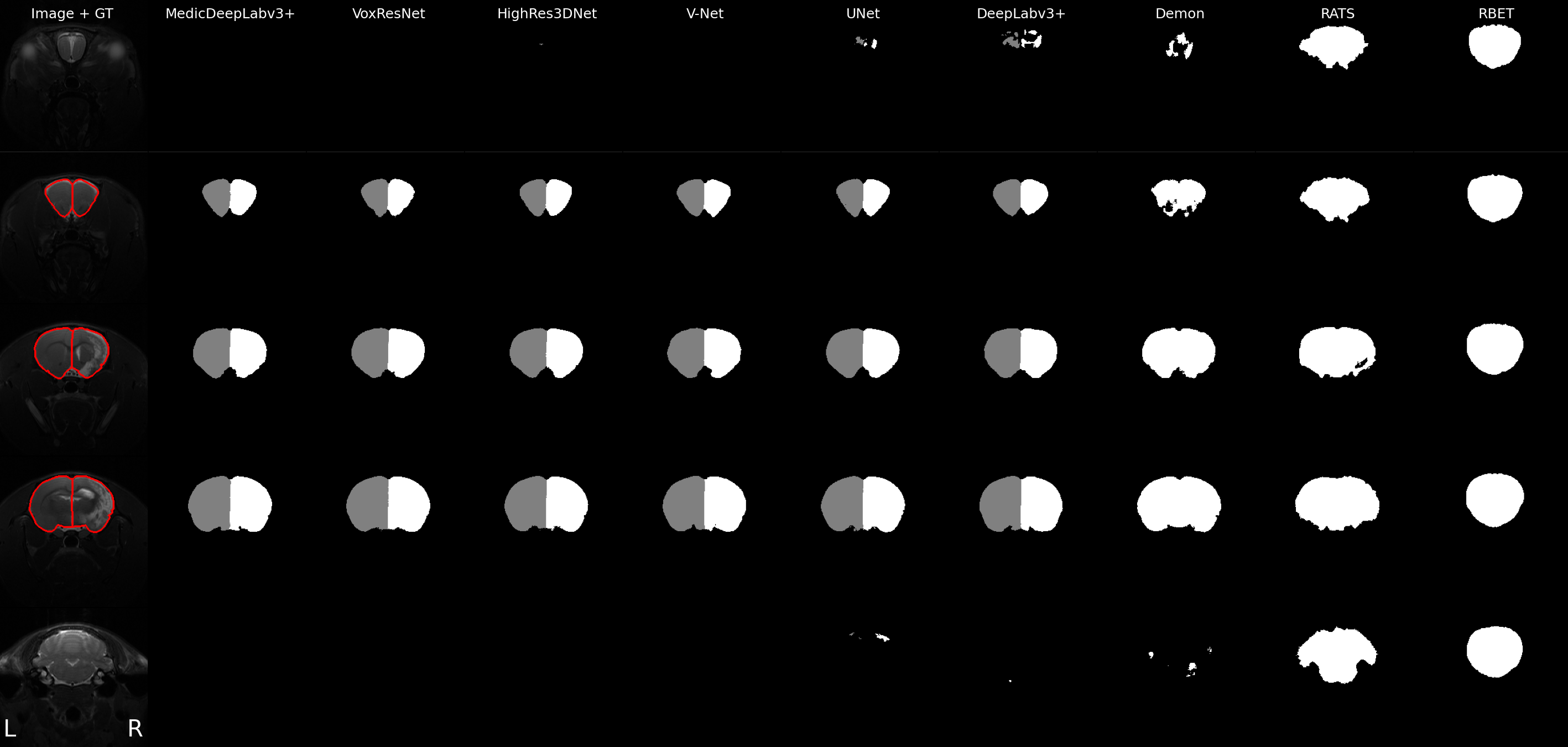}
\caption{T2-weighted image, ground truth and automatic segmentations of a rat from Study 8, D14.}
\label{fig:segmentations}
\end{figure*}

\begin{figure*}[bt!]
\centering
\includegraphics[width=\textwidth]{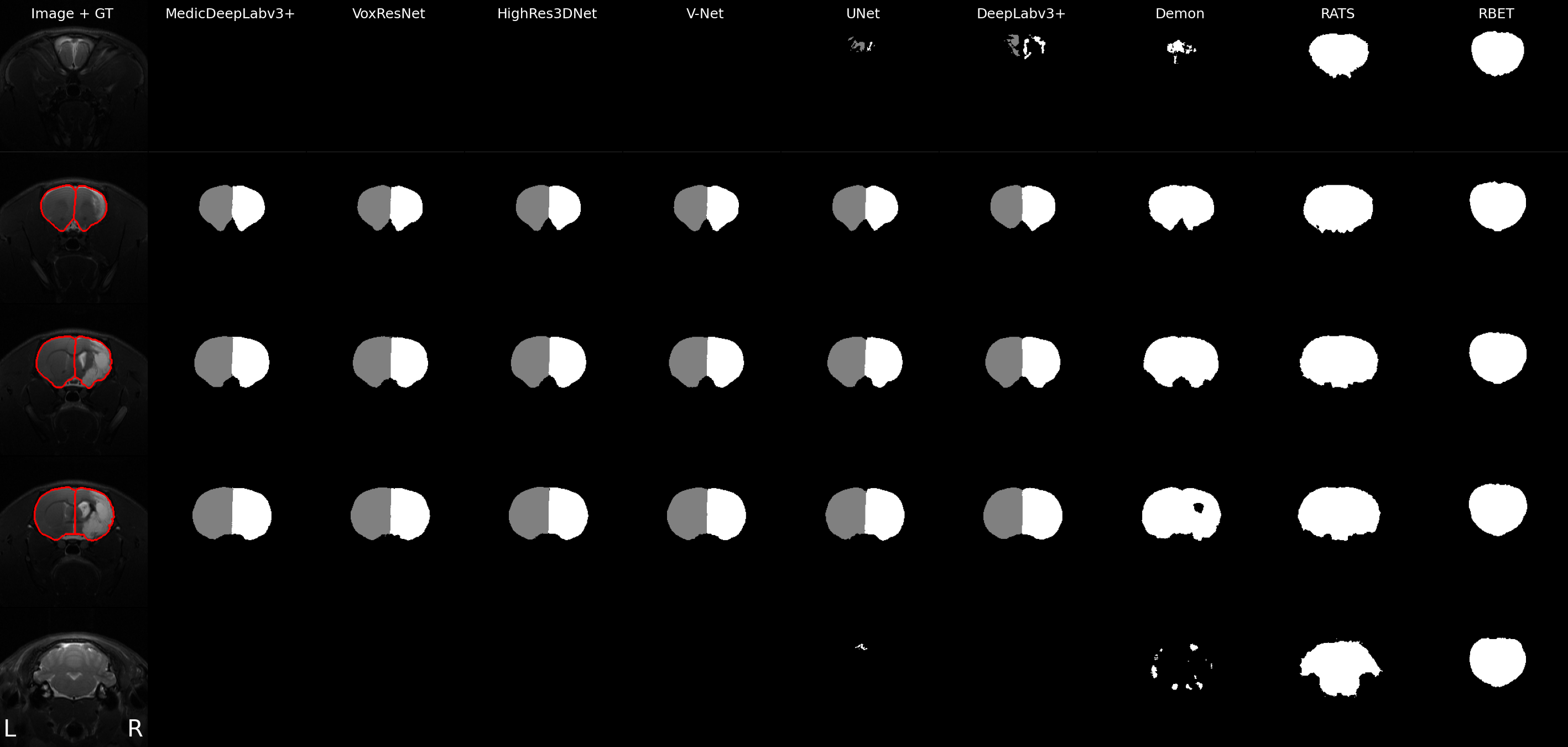}
\caption{T2-weighted image, ground truth and automatic segmentations of a rat from Study 8, D28.}
\label{fig:segmentations}
\end{figure*}

\begin{figure*}[bt!]
\centering
\includegraphics[width=\textwidth]{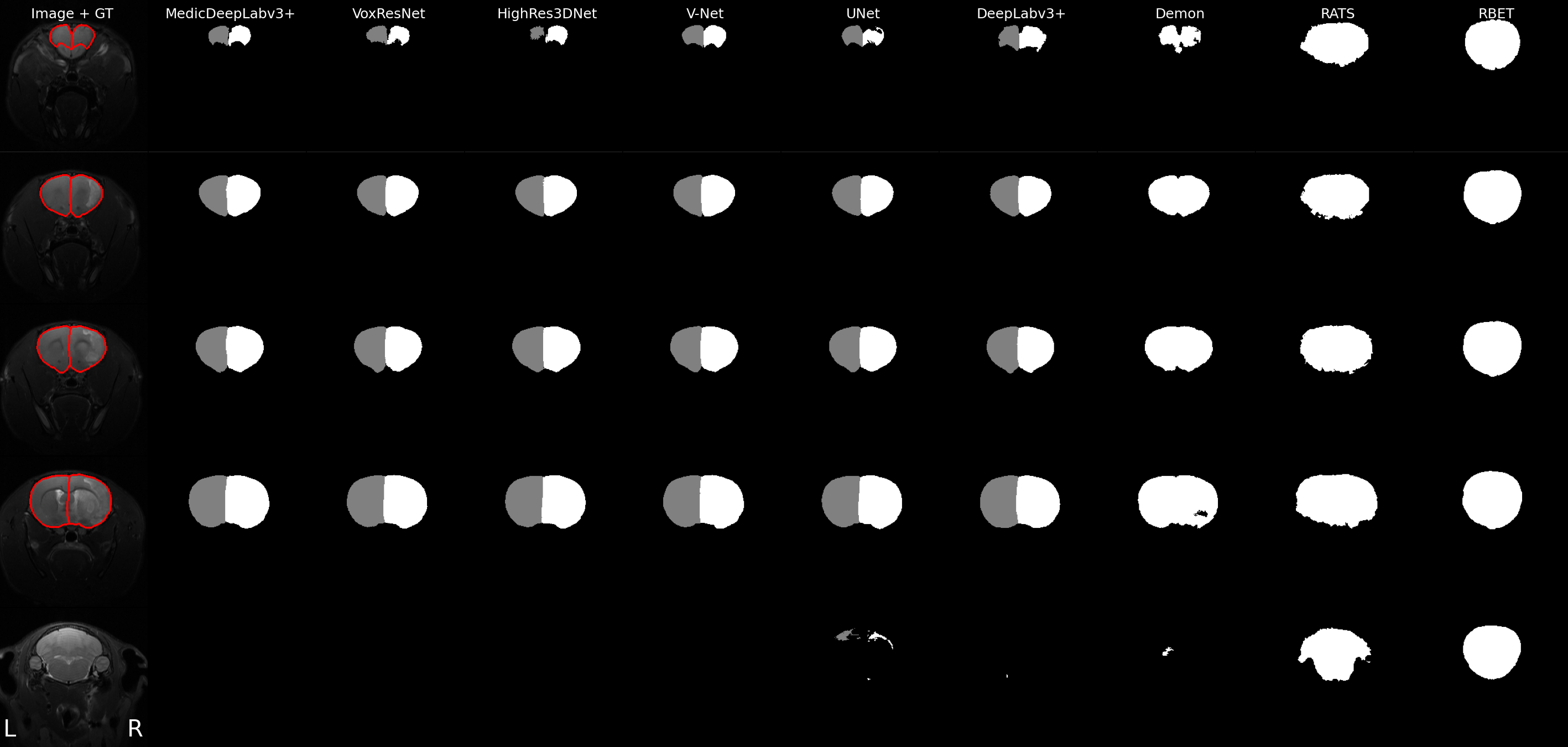}
\caption{T2-weighted image, ground truth and automatic segmentations of a rat from Study 9, 24h.}
\label{fig:segmentations}
\end{figure*}

\begin{figure*}[bt!]
\centering
\includegraphics[width=\textwidth]{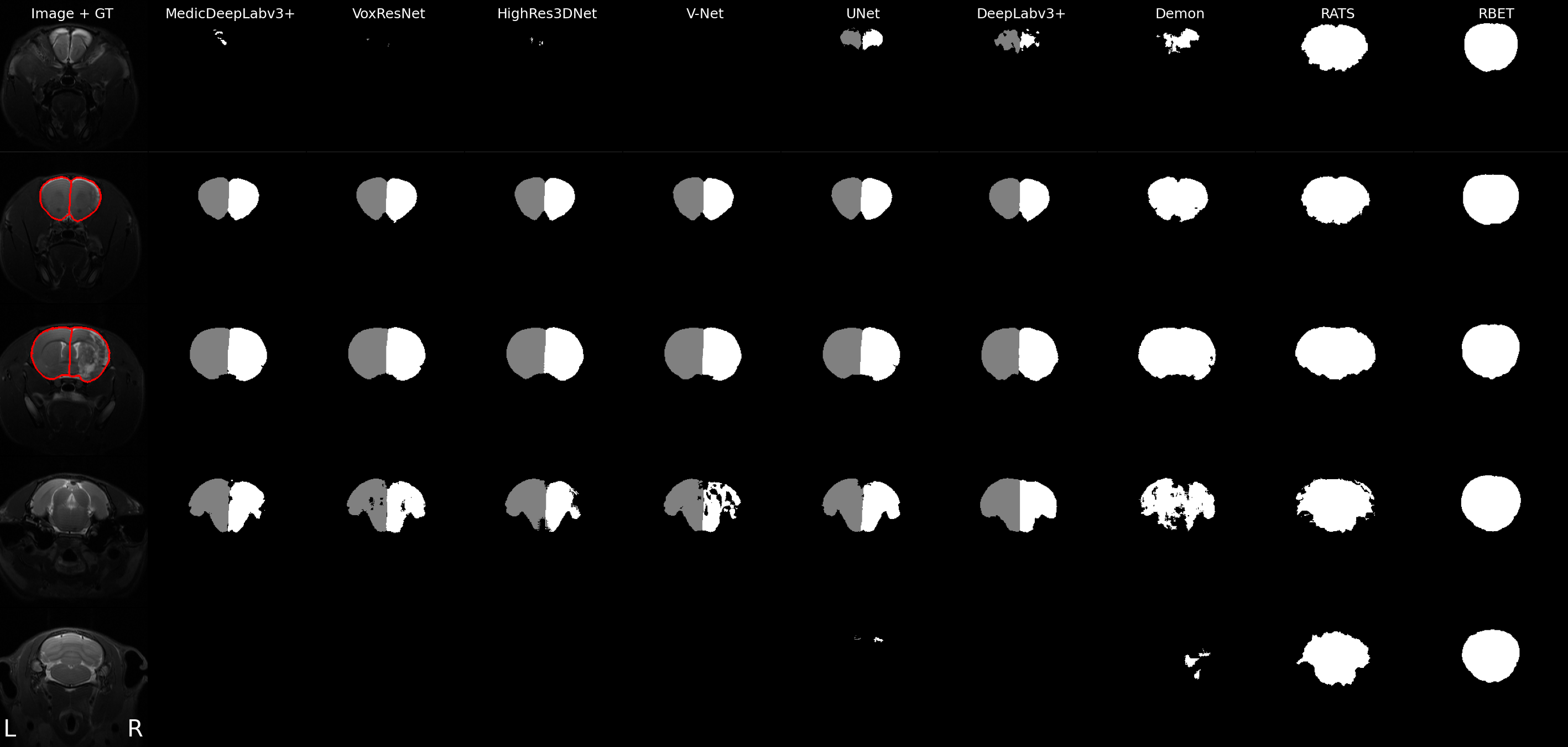}
\caption{T2-weighted image, ground truth and automatic segmentations of a rat from Study 10, D7.}
\label{fig:segmentations}
\end{figure*}

\begin{figure*}[bt!]
\centering
\includegraphics[width=\textwidth]{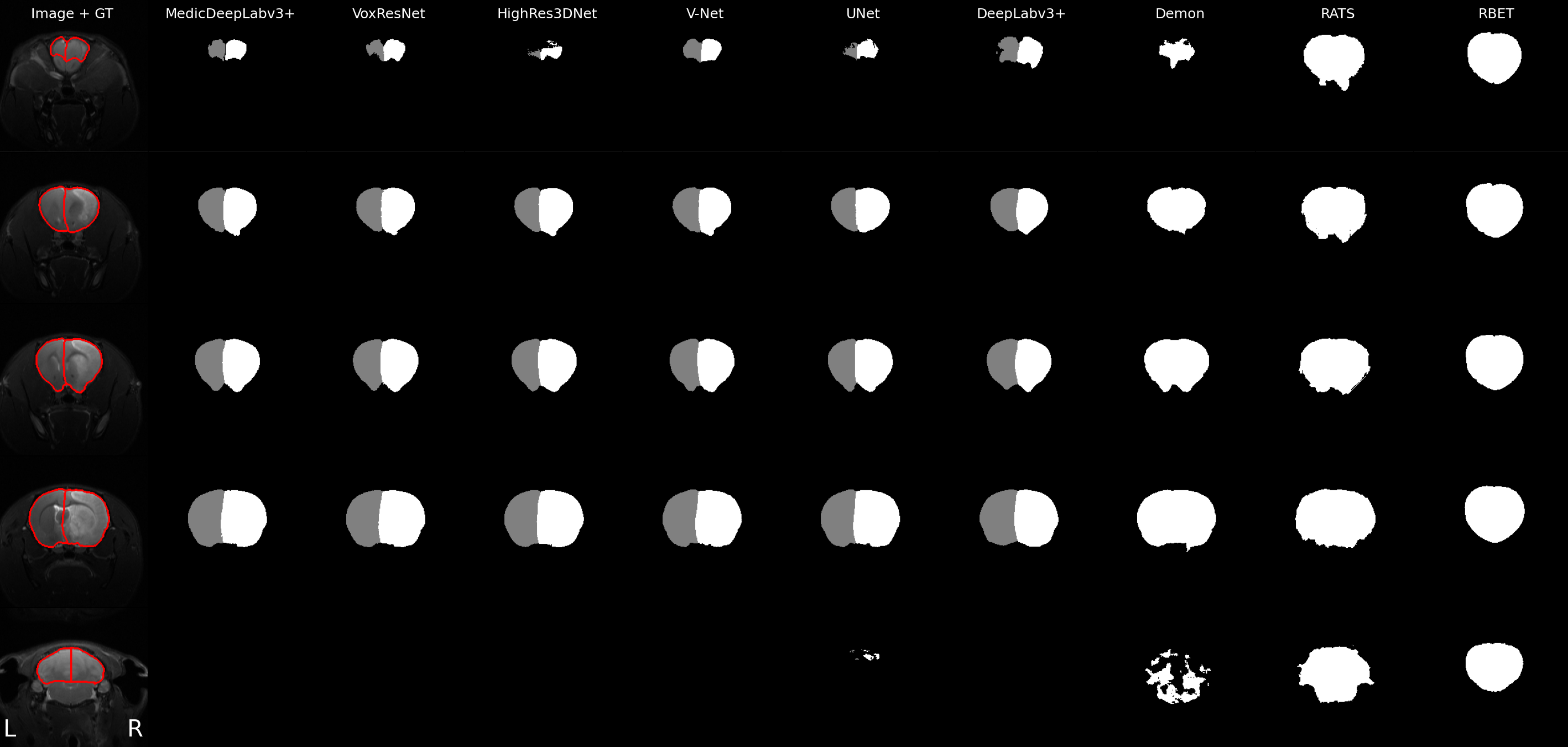}
\caption{T2-weighted image, ground truth and automatic segmentations of a rat from Study 11, 24h.}
\label{fig:segmentations}
\end{figure*}

\clearpage
\section{Supplementary Table}

\begin{table*}[ht]
\begin{center}
\caption{Comparison between multiple versions of MedicDeepLabv3+ with different capacity. Columns: number of initial filters, trainable ConvNet parameters (in millions), optimization time for 300 epochs in our workstation in hours, maximum GPU memory required during training and evaluation, Dice and HD in the brain mask \hl{(mean $\pm$ std)}. Bold: default configuration.}
\label{table:capacitycomparisonsup}
\begin{tabular}{lllllll}
\hline\noalign{\smallskip}
Filters & Parameters & Time (h) & Mem. (train) & Mem. (eval) & Dice & HD \\
\noalign{\smallskip}\hline\noalign{\smallskip}
\textbf{32} & \textbf{79.1M} & \textbf{16.2} & \textbf{8857 MiB} &\textbf{2935 MiB} & \textbf{0.952 \hl{$\pm$ 0.04}} & \hl{1.856 $\pm$ 0.91} \\
28 & 60.7M & 14.4 & 7571 MiB & 2617 MiB & 0.950 \hl{$\pm$ 0.04} & \hl{1.792 $\pm$ 0.95} \\
24 & 44.7M & 12.1 & 6545 MiB & 2319 MiB & 0.950 \hl{$\pm$ 0.04} & \hl{1.759 $\pm$ 1.01} \\
20 & 31.1M & 10.3 & 5619 MiB & 2007 MiB & 0.950 \hl{$\pm$ 0.04} & \hl{1.769 $\pm$ 0.98} \\
16 & 20.0M & 7.8 & 4577 MiB & 1717 MiB & 0.949 \hl{$\pm$ 0.04} & \hl{1.707 $\pm$ 0.97} \\
12 & 11.3M & 6.2 & 3531 MiB & 1421 MiB & 0.948 \hl{$\pm$ 0.04} & \hl{1.747 $\pm$ 1.01} \\
8 & 5.1M & 4.5 & 2503 MiB & 1121 MiB & 0.947 \hl{$\pm$ 0.04} & \textbf{\hl{1.694 $\pm$ 0.93}} \\
\noalign{\smallskip}\hline
\end{tabular}
\end{center}
\end{table*}

\end{document}